\documentclass[prb,aps,twocolumn,floatfix,psfig,epsfig]{revtex4}
\usepackage{graphicx}
\usepackage{amsmath}
\begin{document}
\title{Fractional Quantum Hall Effect in the Second Landau Level: the Importance
of Inter-Composite-Fermion Interaction}
\author{Csaba T\H oke}
\author{Michael R.\ Peterson}
\author{Gun Sang Jeon}
\author{Jainendra K.\ Jain}
\affiliation{Department of Physics, 104 Davey Laboratory, The Pennsylvania State 
University, Pennsylvania, 16802}

\newcommand{\LL}{\ensuremath{\Lambda}L}
\newcommand{\llevel}{\ensuremath{\Lambda}-level}
\newcommand{\llevels}{\ensuremath{\Lambda}-levels}
\newcommand{\myfrac}[2]{\ensuremath{{#1}/{#2}}}

\begin{abstract}
Exact diagonalization of a two-dimensional electron gas in a strong 
magnetic field in the disk geometry shows that there exists a filling factor 
range in the second Landau level where the states significantly differ from 
those in the lowest Landau level.  We show that the difference arises because 
the interaction between composite fermions is not negligible in higher 
Landau levels, as indicated by a substantial  
mixing between composite-fermion Landau-like levels, or \llevels.
We find that the exact ground state is 
well reproduced by composite fermion theory with \llevel\ mixing 
incorporated at the lowest level of approximation.
Using the same variational approach in the spherical geometry we estimate 
the excitation gap at filling $\myfrac{1}{3}$ in the second Landau level (which 
corresponds to 7/3 of experiment).
\end{abstract}

\maketitle

\newcommand{\zeroth}{\ensuremath{0^{\text{th}}}}
\newcommand{\first}{\ensuremath{1^{\text{st}}}}
\newcommand{\second}{\ensuremath{2^{\text{nd}}}}
\newcommand{\third}{\ensuremath{3^{\text{rd}}}}
\newcommand{\Hall}{\ensuremath{\rho_{xy}}}
\newcommand{\Longi}{\ensuremath{\rho_{xx}}}

\section{Introduction}

A two-dimensional electron system (2DES) in a high magnetic field and 
low temperature exhibits a rich structure of phases as a function of the filling
factor $\nu=\rho hc/eB$ ($\rho$ is the electron density and $B$ is the magnetic 
field).
The integral quantum Hall effect (IQHE) \cite{Klitz1} is characterized by 
quantized plateaus in the Hall resistance, where $\Hall=h/ne^2$, and a vanishing
longitudinal magnetoresistance \Longi\ around $\nu\approx n$.
This phenomenon occurs because the 2DES at integer filling factors $\nu=n$ 
is an incompressible liquid with a finite excitation gap.
The fractional quantum Hall effect (FQHE) \cite{Tsui1} refers to plateaus 
at $\Hall=h/fe^2$ in the vicinity of $\nu\approx f$, where $f$ is a fraction.
Our understanding of the physics of the lowest Landau level (LLL) has evolved 
greatly in the last two decades.  The FQHE 
has been explained by the composite fermion 
(CF) theory \cite{Jain1,JainKamilla} as the IQHE of quasiparticles called 
CF's, which are electrons bound to an even number
($2p$) of quantized vortices of the many-body wave function.
CF's feel a reduced magnetic field $B^\ast=B-2p\rho\phi_0$, where 
$\phi_0=hc/e$ is the magnetic flux quantum.
Composite fermions form Landau-like levels in the reduced magnetic field,
which we will call \llevels\ (\LL's).  $\Lambda$-levels are analogous 
to Landau levels (LL's) of electrons at $B^*$.
(The $\Lambda$-levels have been called CF-quasi-Landau levels in the 
literature.  To avoid confusion with the Landau levels of electrons, we 
prefer to use $\Lambda$-levels.  ``Landau level" will refer below exclusively 
to {\em electronic} Landau levels.  Composite fermions can fill many $\Lambda$-levels
within one Landau level.)  When CF's fill an integral number $n$ of \llevels,
an incompressible quantum liquid with a finite excitation gap results,
producing an IQHE of CF's.  Such conditions are 
achieved at electron fillings $\nu=\myfrac{n}{(2pn\pm 1)}$, which are 
precisely the prominently observed fractions.
Further, the microscopic wave functions based on this physics provide an 
excellent account of the actual eigenfunctions.
CF theory with neglected inter-CF interaction thus successfully explains most of 
the observed fractions in the LLL.

This paper is concerned with fractional quantum Hall effect 
in the \second\ LL.  Given that more than 50 fractions have been observed in 
the lowest LL, one might expect a large number of fractions in the second LL 
as well, but FQHE is relatively scarce in the second or higher LL's.
However, with improved experimental conditions (higher mobilities and 
lower temperatures), many fractions have been observed outside of the LLL.  
As seen in the experiments of Xia {\em et al.} \cite{Xia1}, the   
observed fractions are $\nu^{(1)}=\myfrac{1}{3},\myfrac{1}{5},\myfrac{2}{5},
\myfrac{2}{3}$, and $\myfrac{4}{5}$, 
where $\nu^{(1)}$ is the filling factor of the second LL.
(The total filling factor is $\nu=2+\nu^{(1)}$ or $\nu=3+\nu^{(1)}$, with each 
LL contributing two to the filling factor, taking the spin degree of freedom 
into account.)
Gervais {\em et al.} \cite{Gervais1} have also seen evidence for 
FQHE at $\myfrac{1}{5}$ and $\myfrac{4}{5}$ in the {\em third} LL.
Indeed, the decreasing stability of CF formation has been suggested on 
theoretical grounds \cite{Scarola2}, and charge-density wave phases are known to be
dominant in $n\ge 2$ LL's~\cite{Koulakov96}.  The competition between many nearly 
degenerate ground states is illustrated by the observation of the so-called 
re-entrant integral quantum Hall effect: the system goes back and forth between 
the IQHE and the FQHE state several times (see, for example Eisenstein {\em et al.}
\cite{Eisen1} and Xia {\em et al.}~\cite{Xia1}), with the re-entrant IQHE   
interpreted in terms of a localization of a correlated bubble crystal proposed  
theoretically by Koulakov, Fogler, and Shklovskii~\cite{Koulakov96}.

Because the observed fractions are consistent with the expectation 
from a trivial generalization of the CF theory to the second LL, 
it is natural to attempt an explanation in terms of 
composite fermions.  However, the microscopic description of the observed 
fractions in the \second\ LL has not been as successful as in the lowest LL.
As noted by Haldane \cite{Haldane1} and by d'Ambrumenil and Reynolds
\cite{Ambru}, a generalization of the Laughlin wave function for the 
ground state at $\nu=\myfrac{1}{3}$ to the \second\ LL \cite{Mac1} has rather 
poor overlaps with the exact second LL ground state  
(from 0.47 to 0.61 for $4\le N\le 9$ particles), which is to be contrasted with 
near unity overlaps in the lowest LL.
In fact, exact diagonalization studies \cite{Haldane1,Ambru} on small systems  
have been unable to capture conclusively the incompressibility of 
$\nu^{(1)}=\myfrac{1}{3}$, because the system is 
compressible for some particle numbers and incompressible for others, 
and for the incompressible states, the gap varies widely with the number of 
particles \cite{Ambru}.  
The observation of many FQHE states in the filling factor range $2<\nu<4$  
has motivated us to seek a better quantitative understanding 
of the FQHE in the second LL.

Another motivation arises, interestingly, from certain new FQHE states within 
the lowest Landau level.  Pan {\em et al.}\cite{Pan1} have observed several 
fractions ($\myfrac{4}{11}$, $\myfrac{5}{13}$, $\myfrac{6}{17}$, 
$\myfrac{4}{13}$, $\myfrac{5}{17}$, $\myfrac{7}{11}$) that fall outside
the primary Jain sequences $\nu=\myfrac{n}{(2pn\pm 1)}$.  Although these sequences 
exhaust the possible fractions for a system of {\em non}interacting composite 
fermions, the residual interaction between the CF's can generate 
more fractions, in perfect analogy to the 
appearance of the FQHE for electrons because of the Coulomb interaction.  
The new fractions are interpreted in terms of the {\em higher} $\Lambda$-level
{\em fractional} QHE of composite 
fermions~\cite{Jain1,Jain2,Pan1,Park12,Chang1,Lopez1,Goerbig04}.
In particular, $\nu=\myfrac{4}{11}$ is related to the FQHE 
at $\nu^\ast=\myfrac{4}{3}$ of CF's, i.e., $\myfrac{1}{3}$ of CF's in the
second CF \LL. (In sufficiently high magnetic fields the system is fully polarized,
and electrons or composite fermions can be taken to be effectively ``spinless."  
This is equivalent to the limit in which the Zeeman energy is much larger than the 
cyclotron energy.  The filling factor $\nu^{(1)}$ in the second LL corresponds to 
$\nu=1+\nu^{(1)}$ for spinless electrons, but to $\nu=2+\nu^{(1)}$  
for the real ``spinful" electrons of experiment.) 
It was shown by Chang and Jain~\cite{Chang1} that the actual 
4/11 state is extremely well described quantitatively  
by analogy to the 1/3 state in the {\em second} LL, but not to the 1/3 state
in the lowest LL.  That further underscores the need for a better quantitative 
understanding of the FQHE in the second LL.

W\'ojs and Quinn \cite{Wojs} have argued that the difference between the LLL 
and the \second\ LL physics is due to some kind of pairing
in the \second\ LL.  They study the occupation number of various relative 
angular momentum $m$ channels in the exact ground states for $8\le N\le 14$ 
electrons in the spherical geometry and find 
that as the filling factor is varied (tuned by the monopole strength $Q$ 
in the spherical geometry, which corresponds to $2Q$ flux quanta 
penetrating the surface of the sphere), there are peaks in the occupation of 
the $m=5$ channel at certain flux values that they identify with  
$\nu=\myfrac{7}{3},\myfrac{5}{2},\myfrac{8}{3}$.  These findings are taken as 
evidence for pairing of electrons in the $m=5$ channel.  
W\'ojs and Quinn thus assign a qualitatively different physics to FQHE in the 
second LL than that in the lowest LL, as reflected in the fact that, 
in the spherical geometry, their paired state at 1/3 in the second LL 
occurs at the LL degeneracy $N_d=3N-4$ as opposed to $N_d=3N-2$ 
for the 1/3 state in the lowest LL.  (The LL degeneracy is related 
to the monopole strength $Q$ by $N_d=2Q+3$ in the second LL.)
No explicit wave functions have been constructed for the 
conjectured paired states which can be compared to the exact wave functions.
It is noted that 
the likely explanation for the FQHE at $\nu=\myfrac{5}{2}$ \cite{Willett1}
is based on the notion of pairing of composite fermions  
\cite{Greiter91,Morf1,Rezayi,Scarola1} described by a Pfaffian wave function 
of Moore and Read \cite{Moore1}.

Our approach is different.  We attribute the same {\em qualitative} 
physics to the odd-denominator FQHE in the lowest and the second LL's 
(the half filled LL state behaves very differently in the two LL's) 
and argue that the {\em quantitative} differences arise because of  
substantial $\Lambda$-level mixing in the second LL.
The $\Lambda$-level mixing is a signature of the residual interaction between 
CF's.  The negligibility of $\Lambda$L mixing for the lowest 
LL FQHE states is taken to imply that the CF's are weakly 
interacting.  We will see that composite fermions are more strongly interacting
in a range of filling factor in 
the second LL, although not so strongly as to destabilize all FQHE completely. 
The pairing at 5/2 is already recognized as a consequence of a weak 
attractive interaction between composite fermions.

The higher LL FQHE has also been investigated by Goerbig, Lederer and 
Morias-Smith~\cite{Goerbig04b} using the Murthy-Shankar formulation 
of composite fermions~\cite{Murthy03}. They do not consider $\Lambda$L 
mixing in their approach.

We will restrict electrons to the second LL and neglect any LL mixing.
(In practice, we map the second LL problem into an effective lowest 
LL problem, and work within the LLL.)  The $\Lambda$L mixing is treated 
perturbatively by diagonalizing the full Hamiltonian within 
a correlated basis which includes the ``unperturbed" CF ground state as well 
as the ``unperturbed" particle-hole pair excitations across $\Lambda$-levels.  
This is called CF diagonalization.  The resulting ground state incorporates $\Lambda$L 
mixing, and can be improved perturbatively by inclusion of successively higher 
energy CF excitons.   With minimal $\Lambda$L mixing (keeping 
the lowest energy CF exciton),  CF diagonalization produces explicit 
wave functions which, when tested against exact wave function, are found 
to be excellent approximations to the ground states, thus demonstrating that 
$\Lambda$L mixing captures the physics of the second LL FQHE. 
This will be seen to be the case for the entire filling factor range 
where the second LL behaves differently from the lowest LL.  In particular, 
for $\nu^{(1)}=1/3$, the overlap with the exact wave function increases from 
0.71 to 0.95 (for six particles in the disk geometry) upon lowest-order 
$\Lambda$L mixing.

The $\Lambda$L mixing caused by the residual inter-CF interactions 
has been studied through CF diagonalization
previously in various other contexts, for example 
the quantum dot states \cite{Jeon}, the Luttinger liquid at the edge
of the FQHE \cite{Mandal1}, the stability of the FQH liquid state
against CF excitons \cite{Peterson1}, and for obtaining improved variational 
bounds for the FQHE ground state energies in the lowest LL \cite{Mandal2}. 
It has quantitative, and sometimes even qualitative consequences.

The plan of the rest of the paper is as follows.
The next section shows, by comparison of the exact ground states in the 
lowest, the second and the third Landau levels for six electrons on a disk, that there 
exists a filling factor range where they behave differently. 
Sec.~\ref{CFdiag} will outline the method of CF diagonalization,
which incorporates the inter-CF interactions through $\Lambda$L mixing. 
Sec.~\ref{resultssec} will include the effect of $\Lambda$L mixing 
at the lowest order to obtain modified variational ground states in the second 
LL, and find high overlaps, 
between 0.94 and 0.98, with the exact ground state. 
We then proceed to use the same method in the spherical 
geometry to estimate the excitation gap at $\nu^{(1)}=\myfrac{1}{3}$ in Sec.~\ref{gapsection}.

\section{Exact diagonalization}

\subsection{Basics}

The Hamiltonian for the 2DES is
\begin{eqnarray}
\label{Hamilton}
H &=& H_K + H_I + g\mu\mathbf B\cdot\mathbf S;\\
H_K &=& \frac{1}{2m_b}\sum_j\left(\mathbf p_j +\frac{e}{c}\mathbf A(\mathbf r_j)\right)^2,\notag\\
H_I &=& \frac{e^2}{\epsilon}\sum_{i<j}\frac{1}{\left|\mathbf r_{i}-\mathbf r_{j}\right|},\notag
\end{eqnarray}
where $m_b$ is the band mass of the electron and $\epsilon$ is the dielectric 
constant of the host semiconductor.
We use complex coordinates $z=x-iy$ on the plane and a symmetric gauge vector 
potential $\mathbf A=\left(\myfrac{-By}{2},\myfrac{Bx}{2},0\right)$.  
The distance will be measured in units of the magnetic length 
$l_B=\sqrt{\hbar c/eB}$ and interaction energy in units of $e^2/\epsilon l_B$.
The angular momentum $J_z$ commutes with both $H_K$ and $H_I$, and the 
eigenfunctions of $J_z$ and $H_K$ may be written as
\begin{equation}
\label{basisstate}
\psi_{n,l}(z)=\frac{1}{\sqrt{2\pi n!l!}}(a^\dag)^n(b^\dag)^l\exp\left(-\frac{|z|^2}{4}\right),
\end{equation}
where the lowering operators are defined as
\begin{equation}
a=\frac{\alpha+\beta}{\sqrt 2},\;\; b=\frac{\alpha^\dag-\beta^\dag}{\sqrt 2}
\end{equation}
\begin{equation}
\alpha=\frac{z}{2},\;\;\beta=\partial_x+i\partial_y=2\partial_{z^\ast}
\end{equation}
In terms of the raising and lowering operators, 
the kinetic energy part of the Hamiltonian is given by
\begin{equation}
\label{kinetic}
H_K=\left(a^\dag a + \frac{1}{2}\right)\hbar\omega_c,
\end{equation}
with $\omega_c=eB/m_b c$,
and the angular momentum operator is $J_z=b^\dagger b-a^\dagger a$.
The LL index is $n$, and the angular momentum 
is $l-n$.  The Slater determinants of states given in Eq.~(\ref{basisstate}) 
are used as basis vectors when $H_I$ is diagonalized in a
finite subspace of the Hilbert space.

We will neglect LL mixing throughout this paper, which is a good 
approximation when the interaction strength is small compared to the cyclotron 
energy, and also restrict ourselves to states which are maximally spin polarized
(i.e. fully spin polarized in the topmost partially filled Landau level);  
therefore the last term in Eq.~(\ref{Hamilton}) (the Zeeman term) can 
be dropped.  The exact diagonalization will be carried out in each $L$ sector separately,
where $L$ is the total angular momentum of electrons,
because the Coulomb interaction does not mix states with different $L$.
Fixing the total $L$ confines the electrons to a disk (hence the name 
``disk geometry").  The filling factor can be tuned by varying $L$.
The filling factor is not really a well-defined quantity for a finite 
system, and finite systems do not necessarily represent 
a thermodynamic system with a well-defined filling factor.
For certain wave functions it is possible to write a relation between 
$L$ and $\nu$ for a finite $N$. 
Laughlin's wave function~\cite{Laugh} for $\nu=1/(2p+1)$ (where $p$ is integer) 
satisfies the relation  
\begin{equation}
\label{angmomnu}
\frac{1}{2p+1}=\binom{N}{2}\frac{1}{L}.
\end{equation}
We thus know what $L$ value corresponds to $\nu=1/(2p+1)$ in the disk geometry, and it is 
natural to assume that the filling factor decreases monotonically with $L$.

Two-body interactions in the lowest LL are characterized by the Haldane 
pseudopotentials \cite{Haldane1,Haldane2} $V_m$, which are the 
energies of two electrons in relative angular
momentum $m$ state, $|\psi_m\rangle$,
\begin{eqnarray}
\label{pseudopot}
V_m&=&\frac{\langle\psi_m|V(z_1-z_2)|\psi_m\rangle}{\langle\psi_m|\psi_m\rangle}\label{ppdef}\\
&=&\frac{1}{2^{2m+1}m!}\int rdrV(r)r^{2m}e^{-r^2/4}\notag\\
&=&\int qdq\tilde V(q)L_m(q^2)e^{-q^2},\label{laguerre}
\end{eqnarray}
where $V(\mathbf r)=1/r$ is the Coulomb interaction, $\tilde V(q)$ is its Fourier transform,
and $L_m$ is a Laguerre polynomial.
The interaction part of the Hamiltonian can be 
written in terms of the pseudopotentials: 
\begin{equation}
\label{projection}
H_I|\psi\rangle=\sum_{i<j}\sum_m V_m P^{ij}_m|\psi\rangle,
\end{equation}
where $P^{ij}_m$ projects the wave function of the $i,j$th particles  
into the state of relative angular momentum $m$.
As the spatial part of the fermion wave functions is fully 
antisymmetric (full spin polarization is assumed), 
only the odd $m$ channels are filled.

\subsection{Comparing states in different Landau levels}

When studying the state in $n$th LL we assume that the completely 
filled $0,\dots,(n-1)$ LL's are inert, i.e., we diagonalize in the Hilbert subspace
of the $n$th LL.
The existence of LL raising operators in the planar geometry lets us 
map this problem to the lowest LL by using effective pseudopotentials $V^n_m$:
\[
V^n_m=\frac{\langle\psi^n_m|V(z_1-z_2)|\psi^n_m\rangle}{\langle\psi^n_m|\psi^n_m\rangle},
\]
where $|\psi^n_m\rangle=(a^\dag)^n|\psi_m\rangle$.
One can show that
\begin{equation}
\label{effective}
\tilde V^n(q)=\left(L_n\left(q^2/2\right)\right)^2\tilde V(q),
\end{equation}
where $\tilde V^n(q)$ in the 
Fourier transform of the effective interaction for LL $n$.
Then $V^n_m$ can be calculated in a closed form by Eqs.~(\ref{laguerre}) 
and (\ref{effective}) in a straightforward manner.

\begin{figure}[htbp]
\includegraphics[scale=0.65]{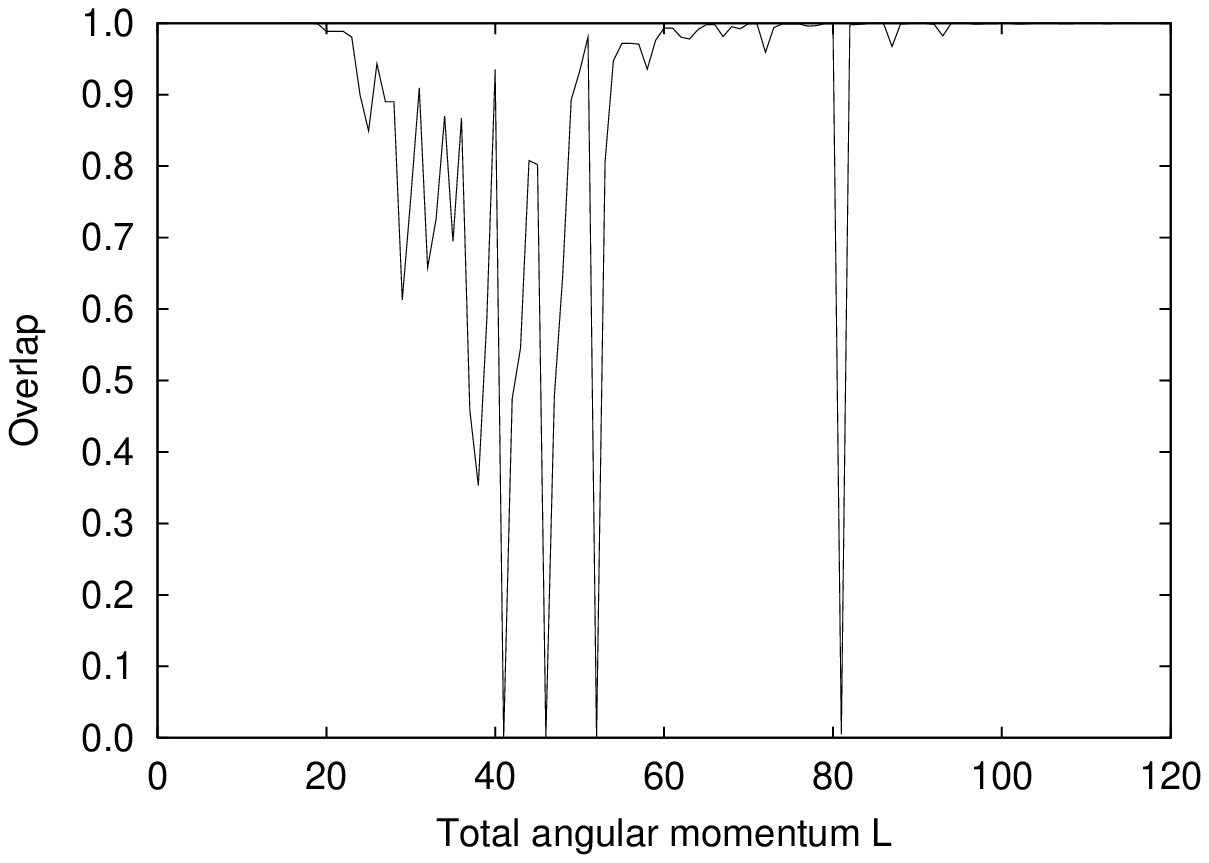}
\includegraphics[scale=0.65]{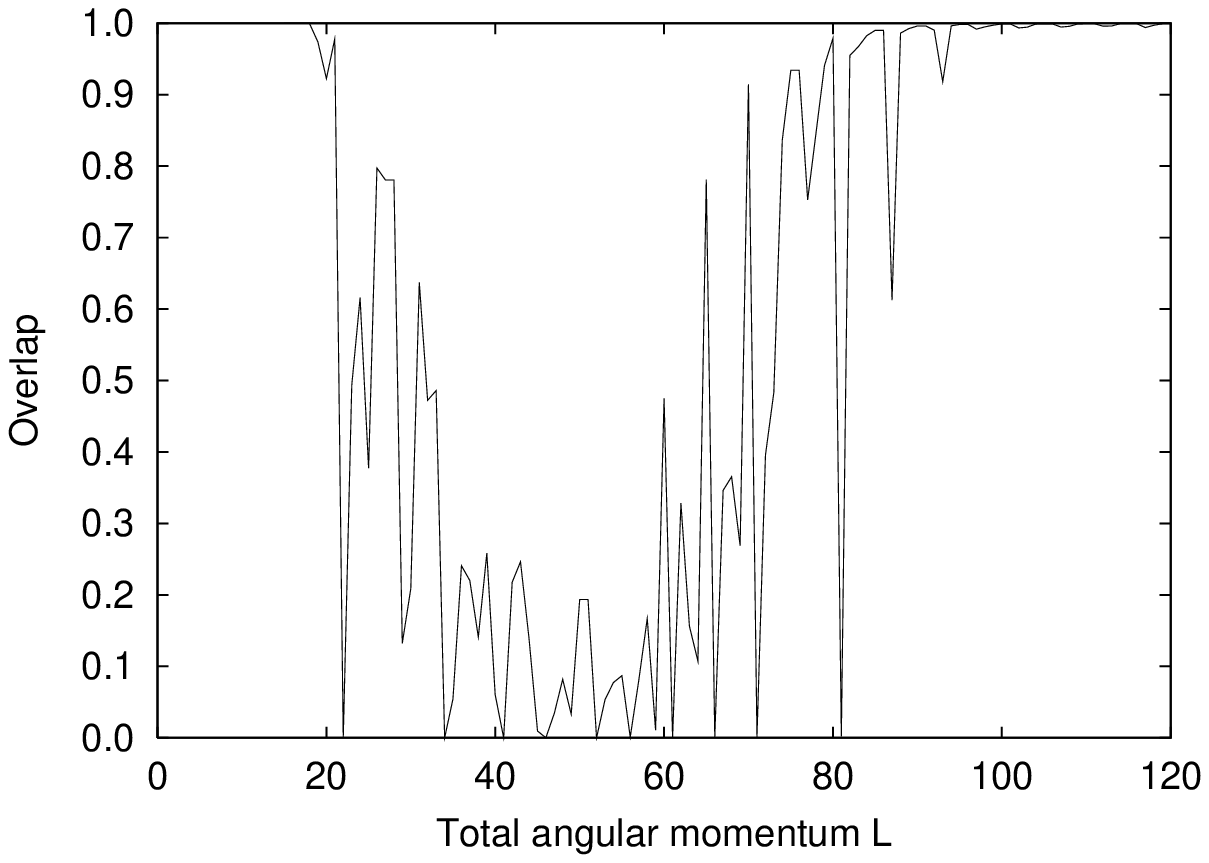}
\caption{\label{differ} Overlap of the LLL ground state 
$|\Psi^{(0)}_{ex}\rangle$ with the \second\ LL ground
state $|\Psi^{(1)}_{ex}\rangle$ (top) and the \third\ LL ground 
state $|\Psi^{(2)}_{ex}\rangle$ (bottom) for $N=6$.
$L$ denotes the total angular momentum.
The overlap $|\langle\Psi^{(0)}_{ex}|\Psi^{(1)}_{ex}\rangle|$ is almost zero 
for $L=41,46,52,81$,
and $|\langle\Psi^{(0)}_{ex}|\Psi^{(2)}_{ex}\rangle|$ nearly vanishes
for $L=22,34,41,46,52,56,61,66,71,81$.
The fillings $\nu=1/3$ and 1/5 occur at $L=45$ and 60, respectively. 
The lines are a guide to the eye.}
\end{figure}

Once we map the higher LL problem with a Coulomb interaction to a LLL problem with 
an effective interaction, we can work within the lowest LL, with different 
interactions simulating different LL's.  This is what we will do.
In the remainder of the article, the 
phrase ``the ground state in the $n$th LL" will really mean  
``lowest LL sibling of the $n$th LL ground state." 
Of course, the actual $n$th level ground state can be obtained from the 
lowest LL wave function by promoting it to the $n$th LL using the 
raising operators.  With that understanding, 
the ground states $|\Psi^{(n)}_{ex}\rangle$ of different LL's can be compared 
by calculating the overlap $\langle\Psi^{(n)}_{ex}|\Psi^{(n')}_{ex}\rangle$.
(Of course, the real $n$th LL ground
state for $n>0$ is orthogonal to the LLL ground state.)

\begin{figure}[htbp]
\includegraphics[scale=0.65]{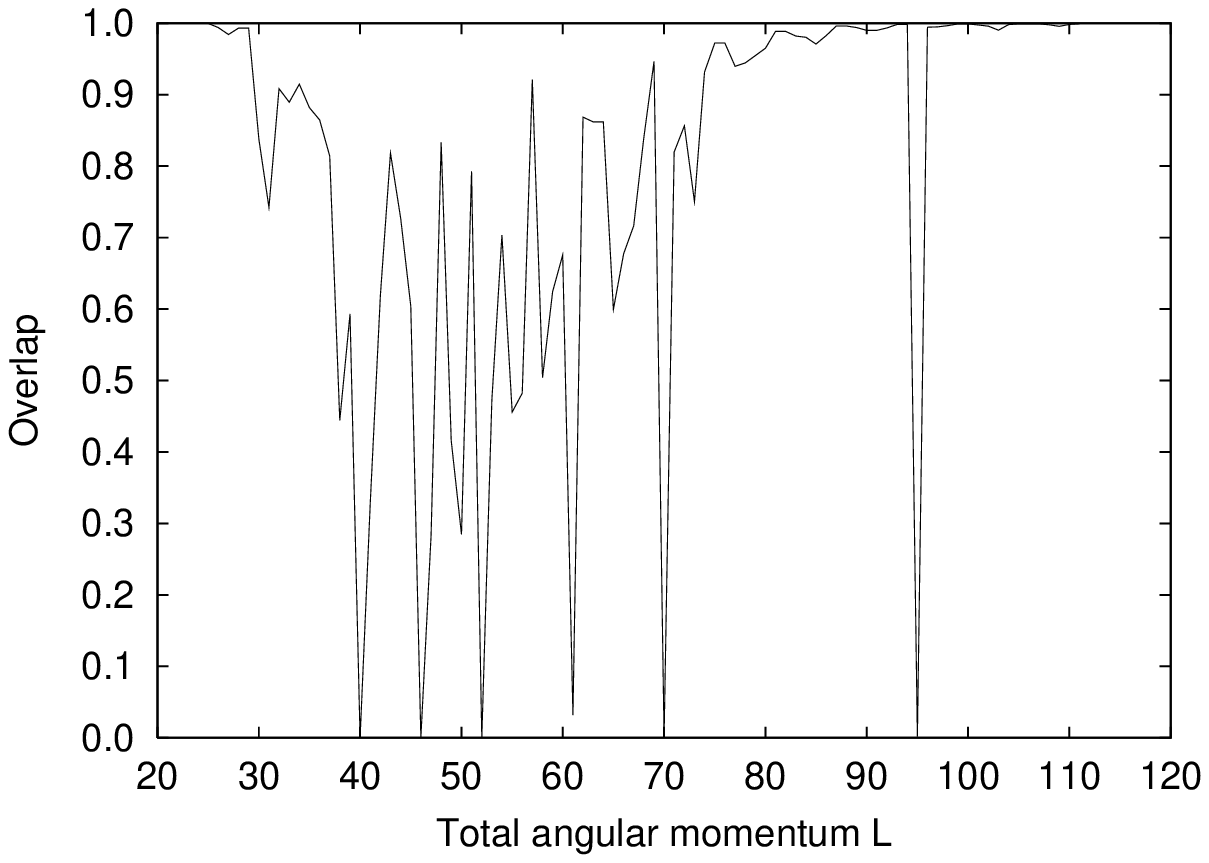}
\includegraphics[scale=0.65]{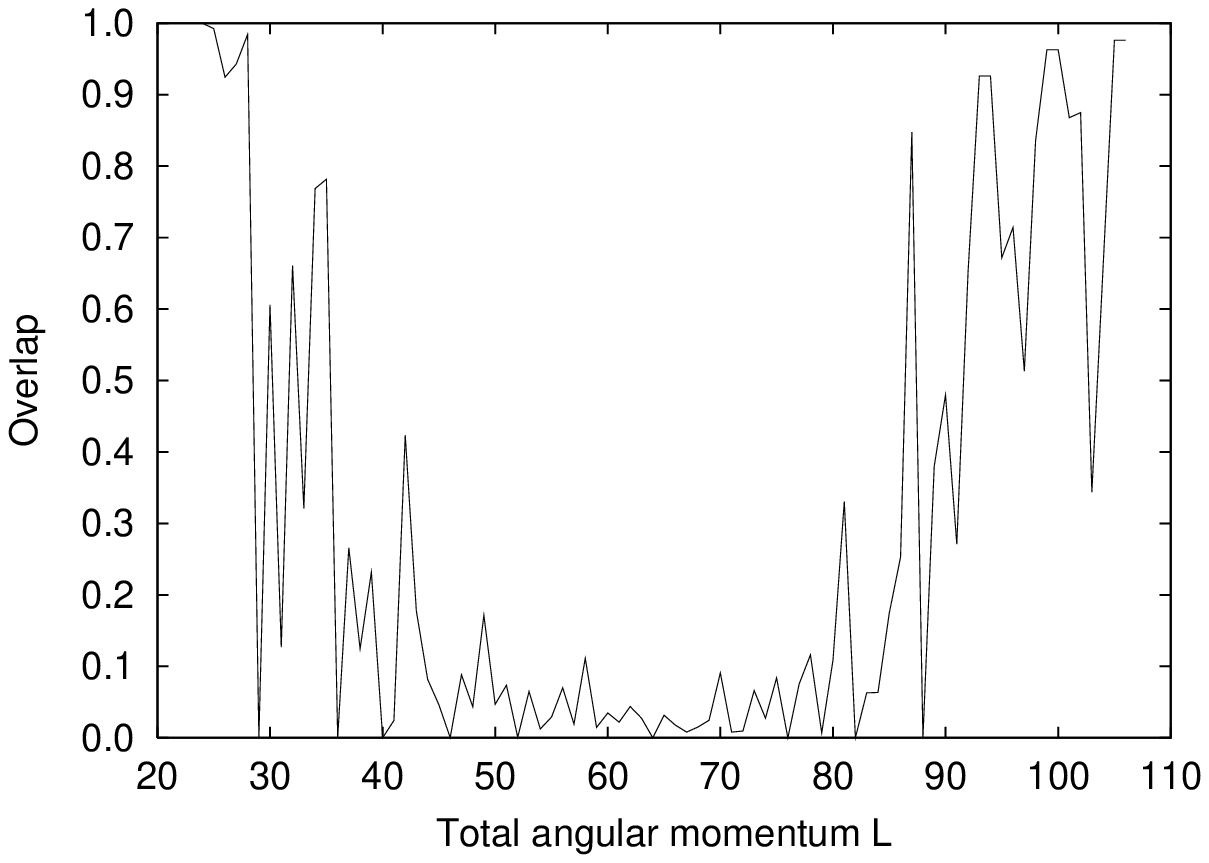}
\caption{\label{differ7} Overlap of the LLL ground 
state $|\Psi^{(0)}_{ex}\rangle$ with the \second\ LL ground
state $|\Psi^{(1)}_{ex}\rangle$ (top) and the \third\ LL ground 
state $|\Psi^{(2)}_{ex}\rangle$ (bottom) for $N=7$.
The overlap $|\langle\Psi^{(0)}_{ex}|\Psi^{(1)}_{ex}\rangle|$ is near 
zero for $L=40,46,52,70,95$ and
quite low ($\approx 0.03$) for $L=61$, where $L$ is the total angular momentum.
The overlap $|\langle\Psi^{(0)}_{ex}|\Psi^{(2)}_{ex}\rangle|$ 
is rather low over a range of $L$.
The fillings $\nu=1/3$ and 1/5 correspond to $L=63$ and 105, respectively. 
The lines are a guide to the eye.}
\end{figure}

\begin{figure}[htbp]
\includegraphics[scale=0.65]{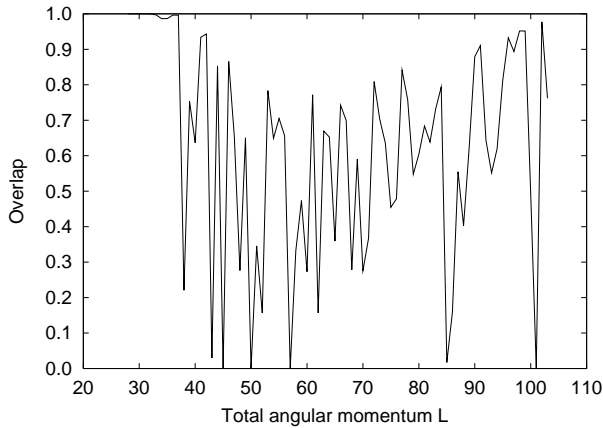}
\caption{\label{differ8} Overlap of the LLL ground state 
$|\Psi^{(0)}_{ex}\rangle$ with the \second\ LL ground
state $|\Psi^{(1)}_{ex}\rangle$ for $N=8$ particles.
The overlap $|\langle\Psi^{(0)}_{ex}|\Psi^{(1)}_{ex}\rangle|$ is almost zero
for $L=45,50,57,85,101$, where $L$ is the total angular momentum.
The fillings $\nu=1/3$ and 1/5 occur at $L=84$ and 140, respectively. 
The lines are a guide to the eye.}
\end{figure}

Fig.~\ref{differ} shows how the overlap between the 
second and lowest (upper panel) and the third and lowest (lower panel) Landau levels 
varies with $L$.
It is apparent that there is an angular momentum range where the
ground states in the lowest and the excited LL's are quite different.
As one considers higher LL's, the range of this region widens.
Similar behavior is found for $N=7$ and $N=8$; see Figs.\ \ref{differ7} and \ref{differ8}.

\begin{figure}[htbp]
\includegraphics[scale=0.5]{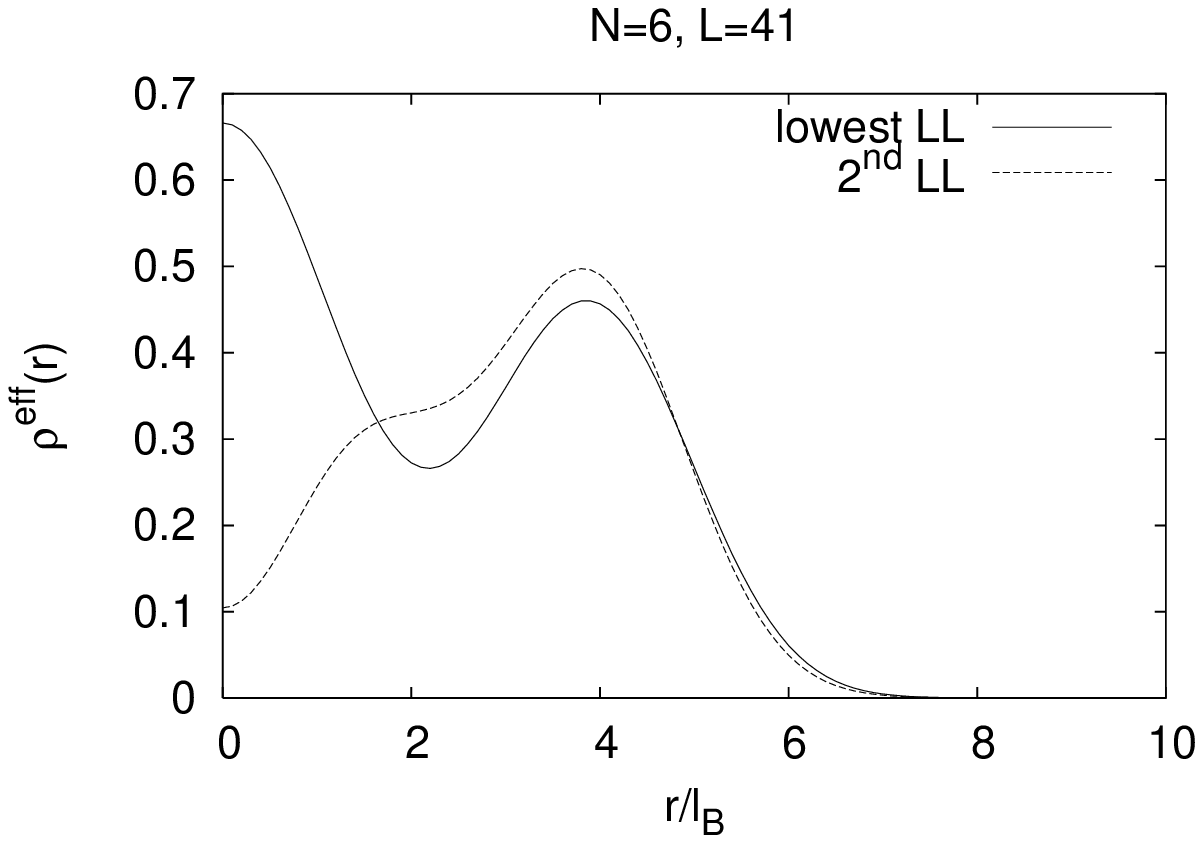}
\includegraphics[scale=0.5]{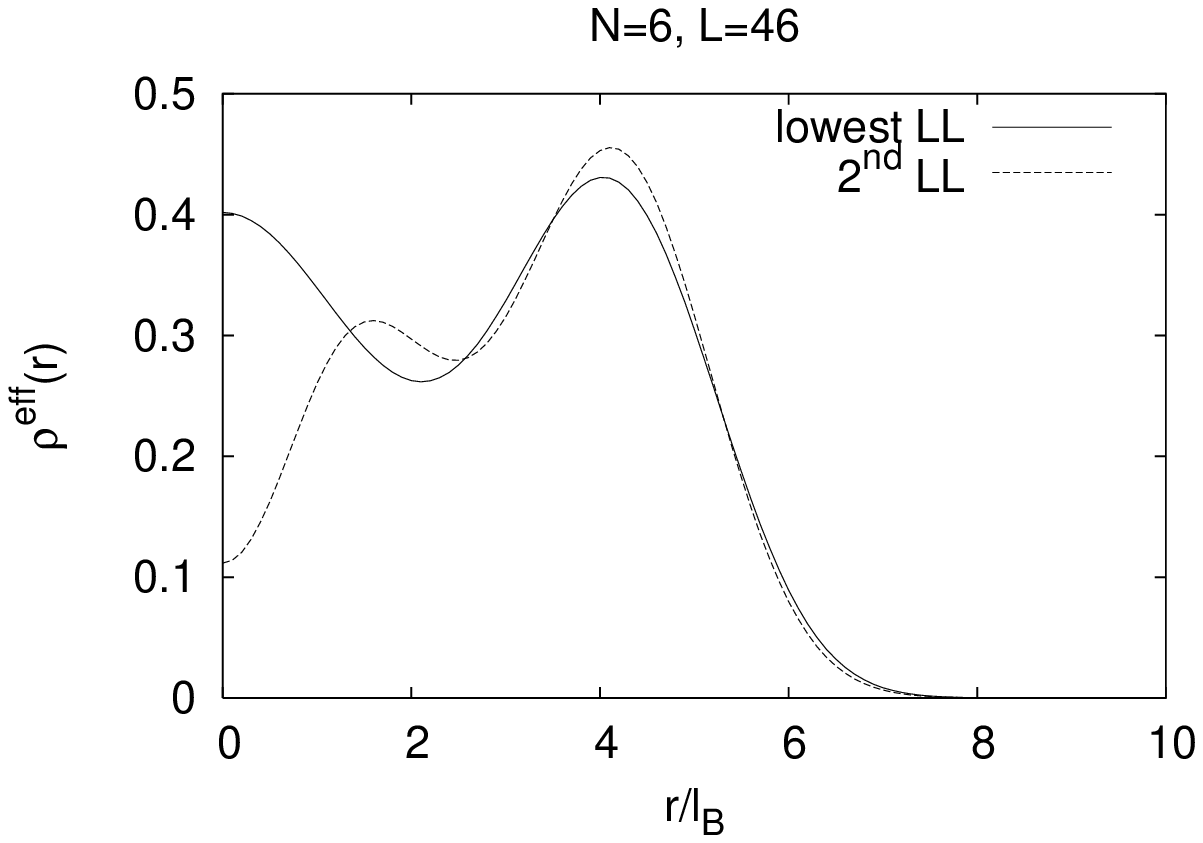}
\includegraphics[scale=0.5]{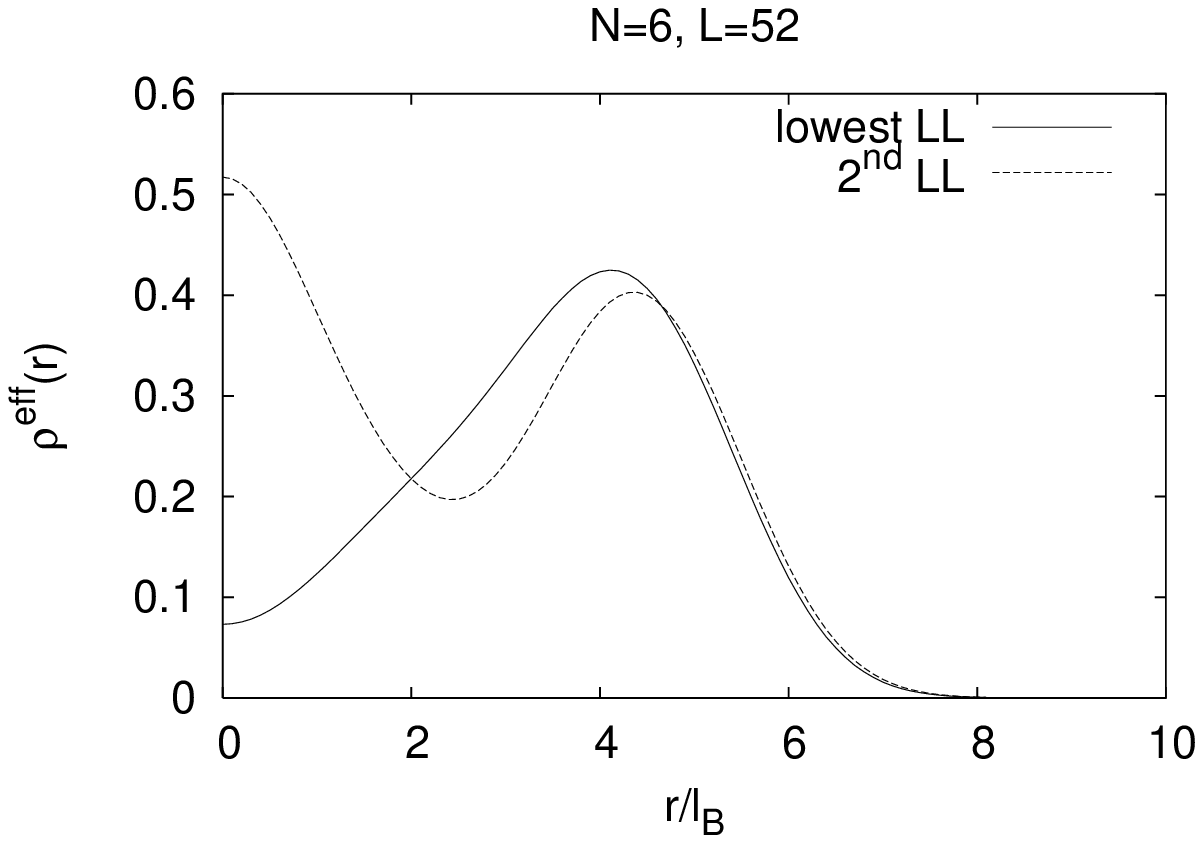}
\includegraphics[scale=0.5]{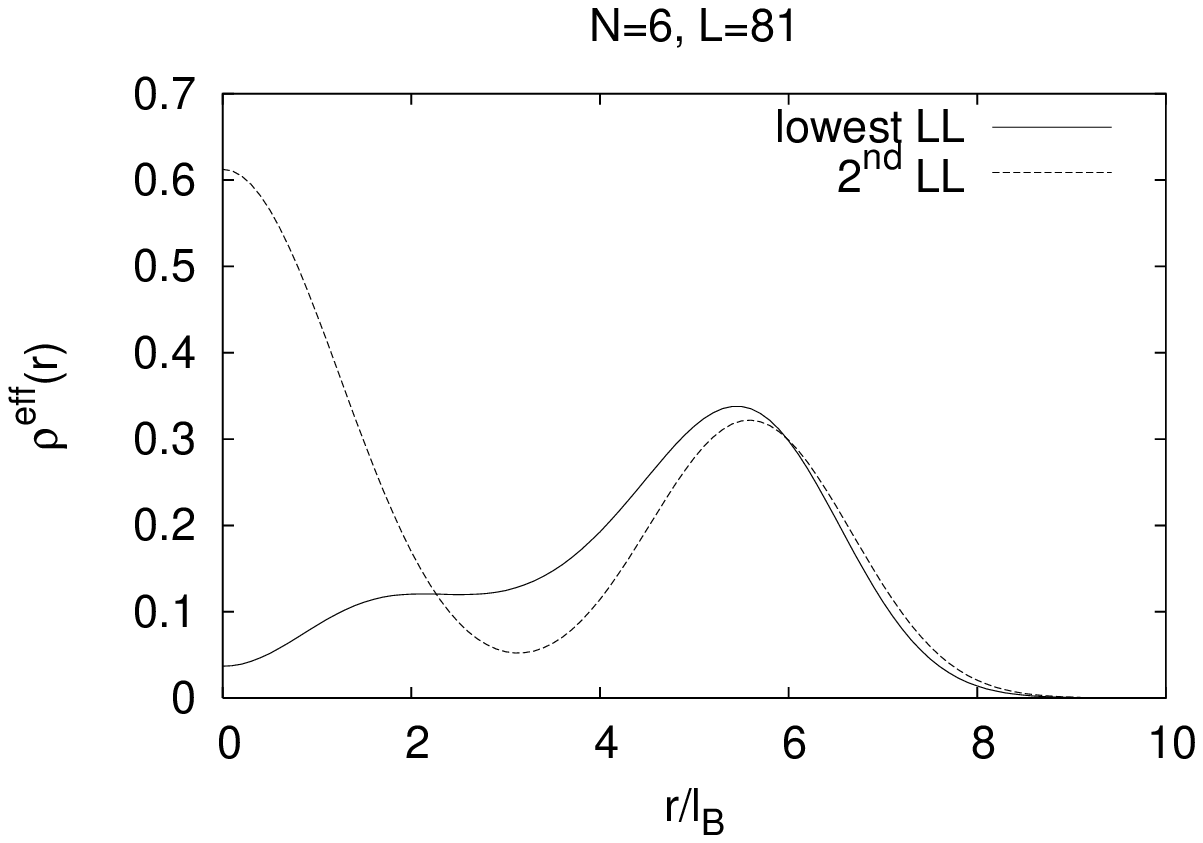}
\caption{\label{low} Comparison of the density profiles $\rho^\text{eff}(r)$ of the 
exact lowest LL and \second\ LL ground states for the angular momentum ($L$) values 
for which the overlap between the two states 
is nearly zero.  The reason for the superscript ``eff" is that the 
density is calculated 
for the lowest LL sibling of the actual ground state.  Note the different behaviors
near the origin. The two Landau levels  prefer crystallites with 
different symmetries at these $L$ values.
The distance is measured in units of $l_B$, the magnetic length.}
\end{figure}

\begin{figure}[htbp]
\includegraphics[scale=0.5]{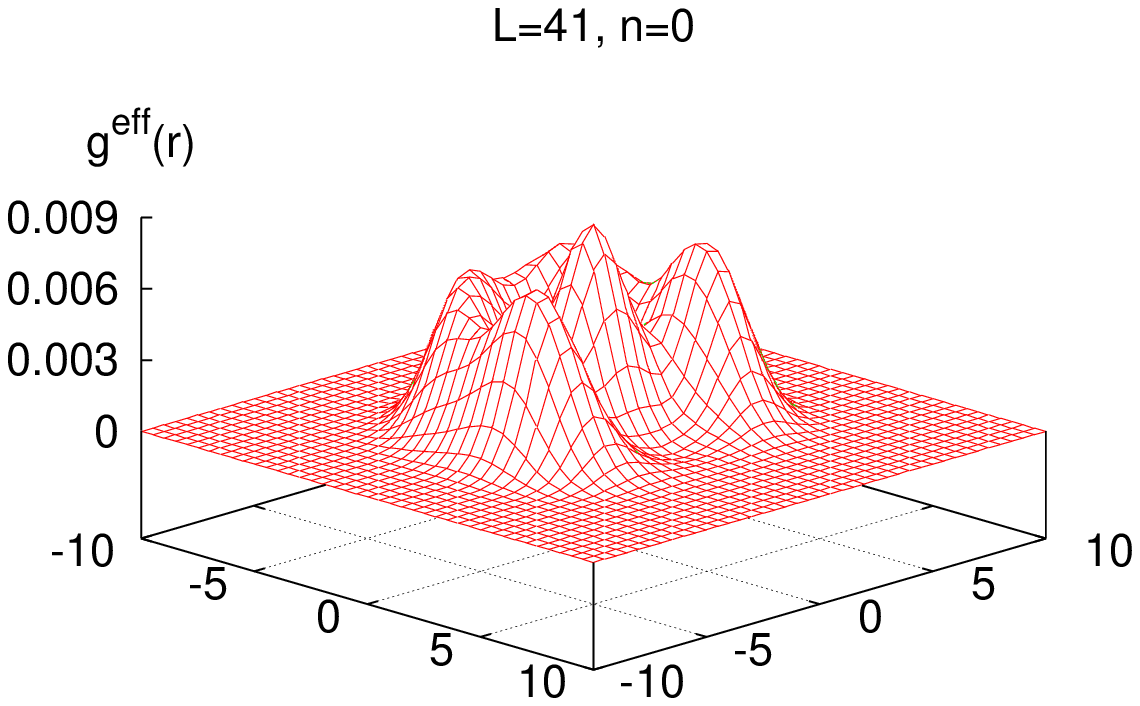}
\includegraphics[scale=0.5]{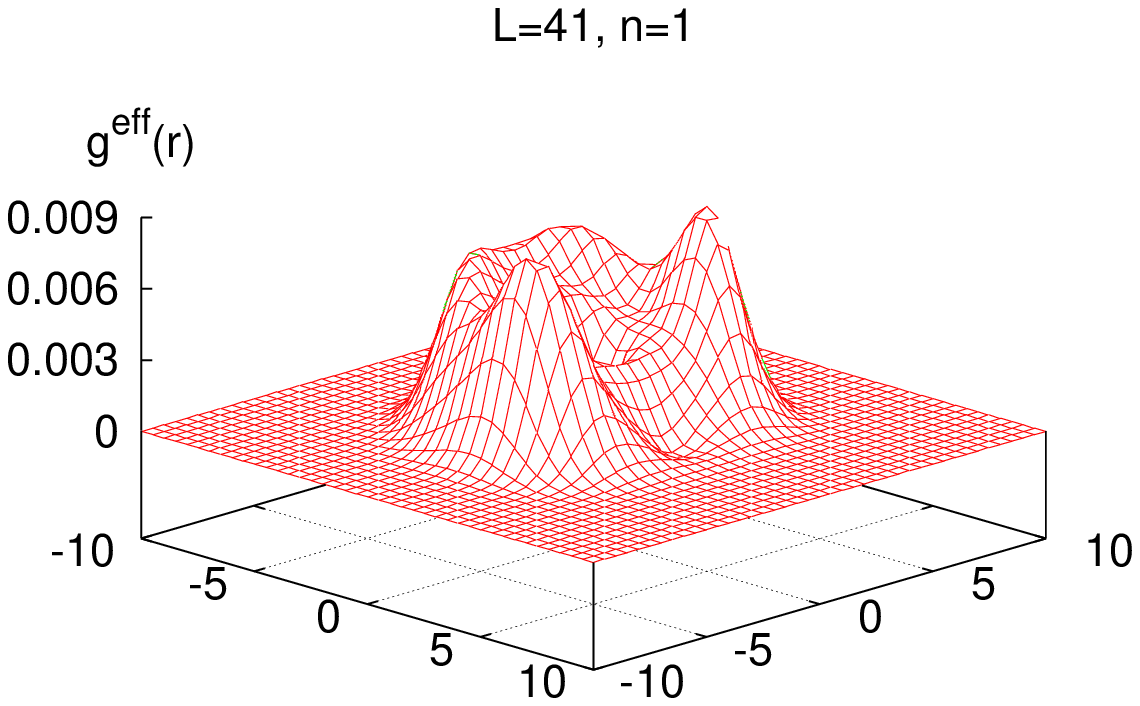}
\includegraphics[scale=0.5]{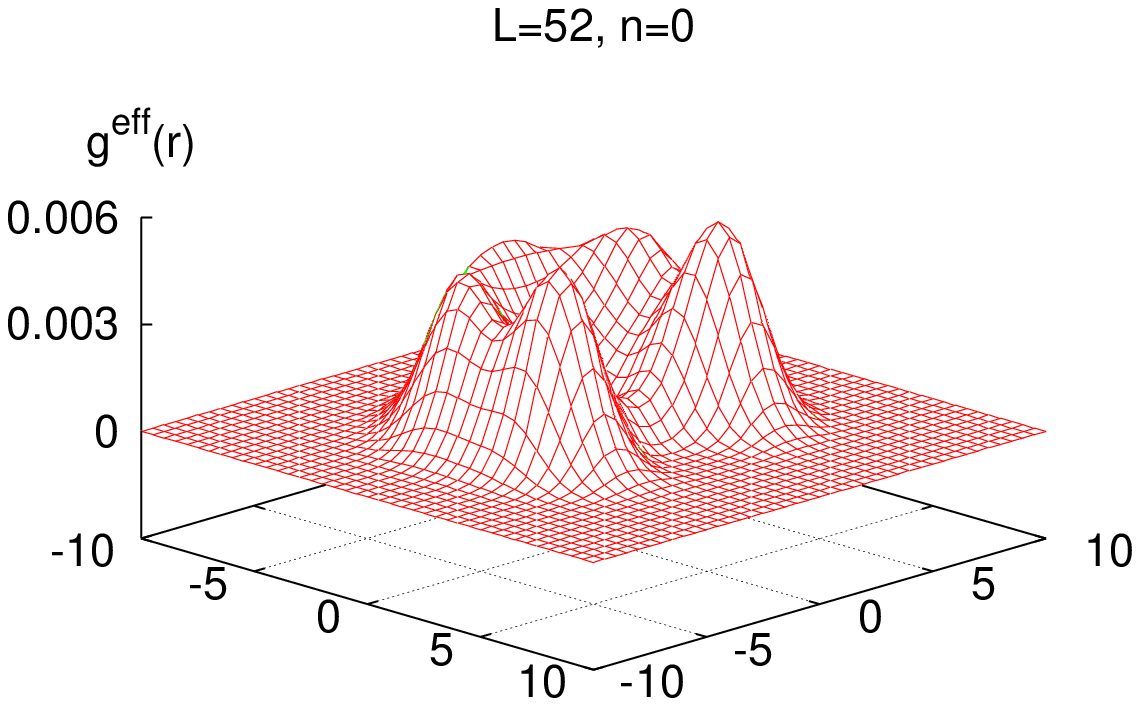}
\includegraphics[scale=0.5]{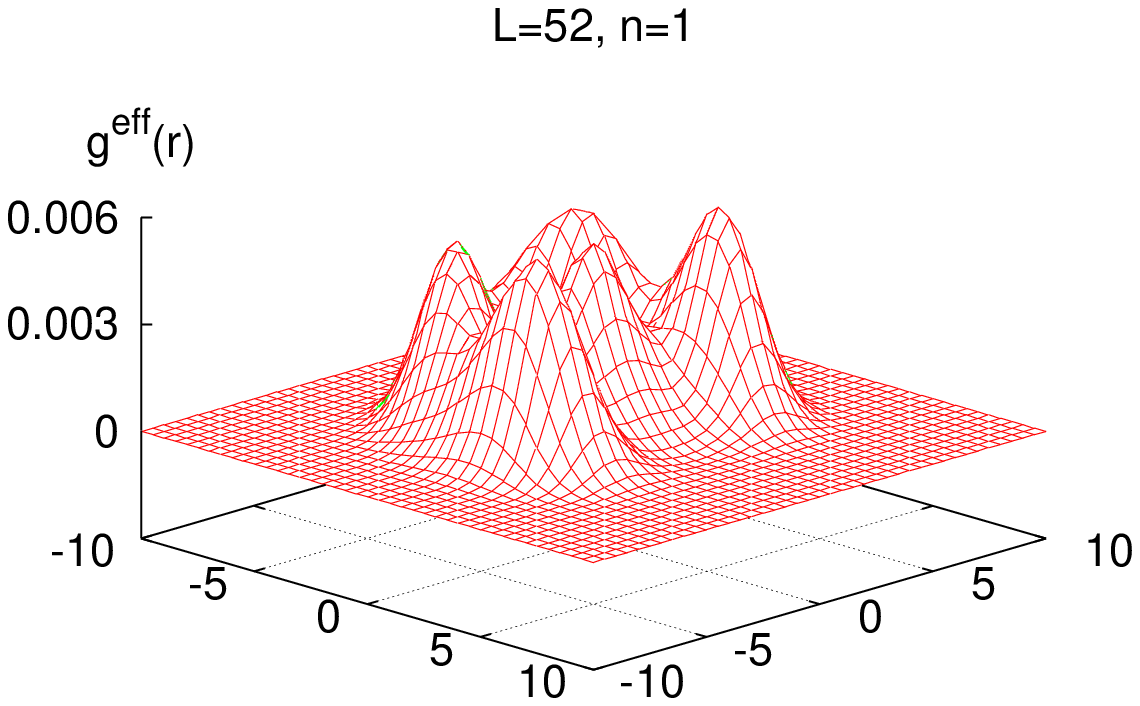}
\caption{\label{pairs}
The ``effective'' pair correlation function $g^\text{eff}(\textbf{r})$ of the ground
state for total angular momenta 
$L=41$ and $L=52$ in the lowest ($n=0$) and the second ($n=1$) Landau levels
The distance $\textbf{r}$ is measured from the center of the disk in units of 
the magnetic length $l_B$.
The reason for the superscript ``eff" is that the pair correlation function 
is calculated for the lowest LL sibling of the actual higher LL ground state.
Similar differences were found for $L=46$ and $L=81$.  }
\end{figure}

There are angular momenta where the ground states in different 
Landau levels are nearly orthogonal.  For example,  for $N=6$
$|\langle\Psi^{(1)}_{ex}|\Psi^{(0)}_{ex}\rangle|=0.00024,0.005$ for $L=41,81$, 
respectively, and is 0 within numerical accuracy for $L=46,52$.
This near orthogonality is most unexpected.  We have found that, in 
most cases, the following simple explanation can be given for it. 
As known from quantum dot studies \cite{quantumdots,Jeon},
two-dimensional few electron systems
in a strong transverse magnetic field have a tendency to form crystallites.
Crystallite structures with $N=6$ fall into two classes of symmetry: 
the (1,5) crystallite resembles a rotating molecule with one particle at the
center and five in a ring; the (0,6) crystallite has all particles on a ring.
Although the density profile differentiates between the two groups,
the pair correlation function contains more information about the crystallite structure.
For $N=6$ and $L=41,46,52,81$ the symmetry of the crystallites in
the two LL's differ, as suggested by the ``effective'' density 
profiles $\rho^\text{eff}(r)$ in Fig.~\ref{low} and demonstrated by the 
pair correlation functions $g^\text{eff}(\textbf{r})$ in Fig.~\ref{pairs}.
(The superscript ``eff" reminds us that 
the density or the pair correlation function are for the lowest LL 
representation of the $n$th LL states.  The ``real" density or the pair correlation 
function in the $n$th LL can be obtained by elevating the wave 
functions to the $n$th LL with the help of the raising operators.)
For $L=41$ and $L=46$ the crystallite in the LLL has $(1,5)$ symmetry while 
the one in the \second\ LL has $(0,6)$ symmetry;
for $L=52$ and $L=81$ the opposite holds.
Although the majority of cases of almost zero overlap can be
understood this way, there are exceptions ($N=7,L=95$ and $N=8,L=101$).
We have not explored this issue further.

Outside a range of $L$, the higher LL physics is similar to that in the 
lowest LL.  It was shown earlier \cite{Ambru,Mac2} that for $\nu\leq 1/5$
the lowest two LL's should show similar behavior.  
The difference at $N=6,L=81$ seems to violate this rule, but the system 
is so small that this exception can be attributed to finite size effects.
We have found no such exception for 
$N=7$ in the $L$ range that we have investigated below 
$\nu=1/5$ ($105< L\le 111$); for $N\ge 8$ the total angular 
momentum corresponding to $\nu=\myfrac{1}{5}$ could not be reached.

\section{CF diagonalization}
\label{CFdiag}

The idea of composite fermion diagonalization is conceptually quite straightforward.
Instead of diagonalizing in the full LLL basis space, as is done 
in exact diagonalization studies, we diagonalize the full Hamiltonian 
in a restricted basis 
of correlated states produced by the CF theory.  The correlated 
basis is constructed as follows.  The CF theory \cite{Jain1,JainKamilla} 
maps interacting electrons at $L$ into non-interacting fermions 
at $L^\ast$ (taking, for concreteness, the disk geometry),
\begin{equation}
L^\ast=L-pN(N-1).
\end{equation}
We begin with the Slater determinant states $\Phi^{L^*}_\alpha$, which 
are some low kinetic energy states at $L^*$ (chosen as explained below), 
and construct Jain's wave functions as:
\begin{equation}
\label{mapping}
\Psi^L_\alpha=\mathcal{P}_{LLL}\Phi^{2p}_1\Phi^{L^\ast}_\alpha,\\
\end{equation}
where $\Phi_1$ is the wave function at $\nu=1$ and $\mathcal{P}_{LLL}$ 
projects the state to the lowest LL.
In the planar geometry,
\begin{equation}
\Phi^{2p}_1=\prod_{j<k}(z_j-z_k)^{2p}\label{Jdisk}.
\end{equation}
(The form for $\Phi_1$ in the spherical geometry is given later.)
The mapping in Eq.~(\ref{mapping}) amounts to a description of the 2DES in 
terms of composite fermions, consisting of the bound state of
an electron with $2p$ quantized vortices of the 2DES wave function. 
$^{2p}\text{CF}$ denotes a CF with flavor $2p$.
The LL's at $L^*$ become $\Lambda$L's of composite fermions 
under the above mapping, and the cyclotron energy at $L^*$ becomes 
the effective cyclotron energy of CF's. 
While the diagonalization in the basis defined by $\Psi^L_\alpha$ is in 
principle straightforward, its actual implementation requires several 
technically challenging steps which have all been demonstrated previously. 
The lowest LL projection will be evaluated following Jain and Kamilla\cite{JainKamilla}.
The states $\Psi^L_\alpha$ are not necessarily 
orthogonal; their orthogonalization and the diagonalization of the Hamiltonian 
matrix requires the evaluation of large dimensional integrals.  Mandal and 
Jain\cite{Mandal1} have demonstrated how this can be carried through in a Monte 
Carlo scheme.  We note that Monte Carlo is used only to evaluate integrals.

\begin{figure}[htbp]
\includegraphics[scale=0.28]{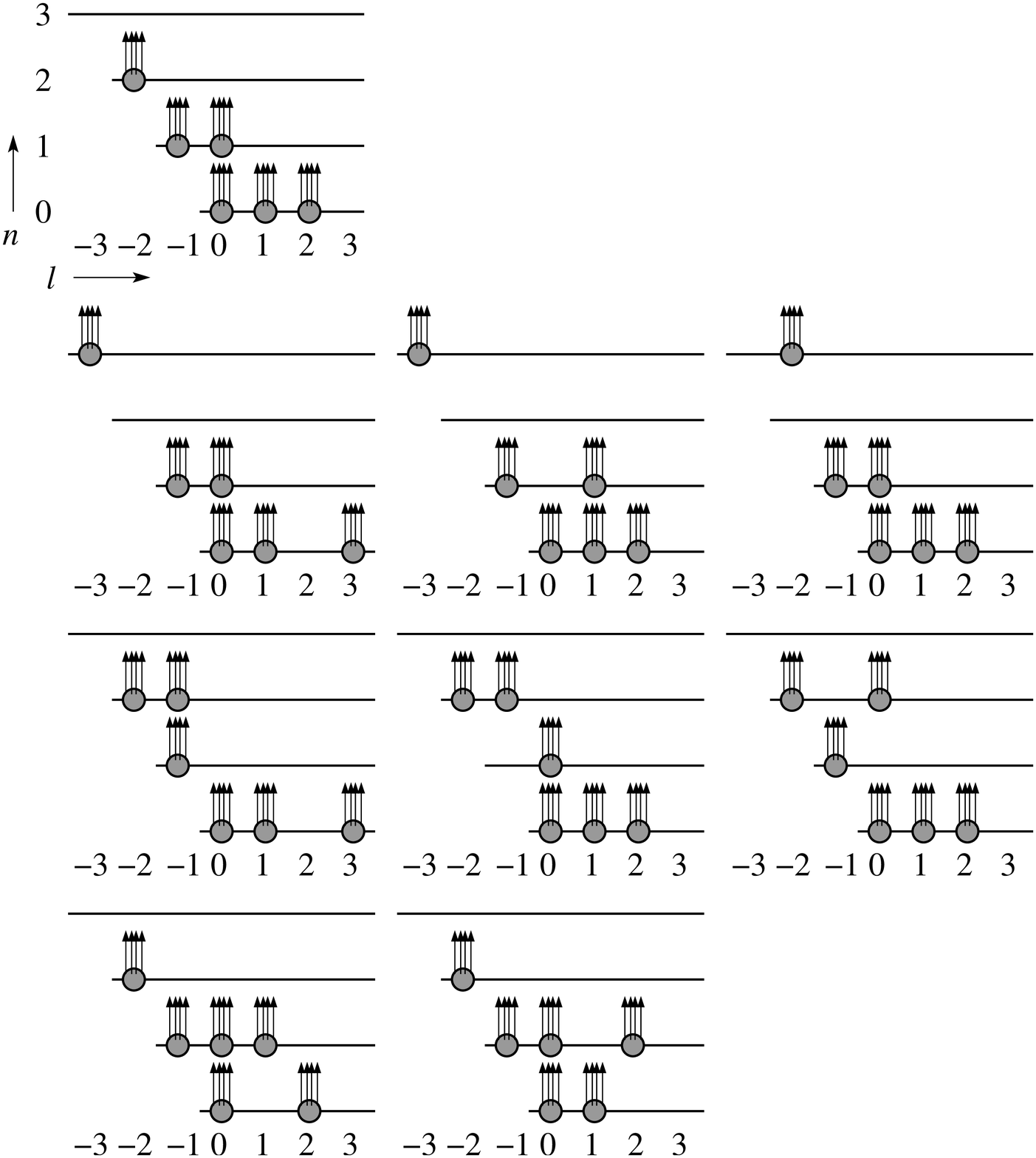}
\caption{\label{L60}  The CF basis states (schematically) for $N=6$ electron 
system at $L=60$, 
which maps into $L^*=L-2N(N-1)=0$ of $^4$CF's (composite fermions carrying 
four quantized vortices).  The y-axis shows the $\Lambda$L 
index, and the x-axis the angular momentum index.
The top row shows the state with the smallest effective cyclotron energy; 
there is only one state at that energy.
The remaining figures show all the linearly independent configurations  
with one additional unit of kinetic energy.
(There are a total of 10 distinct electron states in the $J=1$ basis at 
$L^*$, which produce 9 linearly independent CF basis states.)}
\end{figure}

Next we come to the choice of the basis $\{\Phi^{L^\ast}_\alpha\}$ at $L^*$.
Electrons at $L^*$ will in general occupy several Landau levels, which 
map into $\Lambda$L's of composite fermions through the above 
construction.  At the simplest level (referred to as the zeroth order approximation), 
these basis states are chosen to be the 
distinct degenerate kinetic energy {\em ground} states of non-interacting fermions 
at $L^*$.  The $\Lambda$L mixing corresponds to LL mixing of fermions at $L^*$, 
and can be incorporated into theory by considering a larger basis:
\begin{eqnarray}
\label{L*basis}
[\{\Phi^{(0)}_{\alpha}\},\{\Phi_{\beta}^{(1)}\},\{\Phi^{(2)}_\gamma\},\ldots,
\{\Phi^{(J)}_\zeta\}]\;,
\end{eqnarray}
where $\{\Phi_{\eta}^{(J)}\}$ represents all $\eta$ basis wave functions at $L^*$
with kinetic energy $J$ (in units of the cyclotron energy)
relative to the ground state kinetic energy.
The corresponding CF basis is given by   
\begin{eqnarray}
\label{basis}
[\{\Psi^{(0)}_{\alpha}\},\{\Psi_{\beta}^{(1)}\},\{\Psi^{(2)}_\gamma\},\ldots,
\{\Psi^{(J)}_\zeta\}]\;,
\end{eqnarray}
with $\Psi^{(j)}_{\eta}=P_{LLL}\Phi^{2p}_1\Phi_{\eta}^{(j)}$.
A diagonalization in this basis incorporates the effect of $\Lambda$L mixing
perturbatively.  Fig.~\ref{L60} explains the physics pictorially.
We will denote by $\chi^{(J)}$ and $E^{(J)}$ 
the ground state wave function and the ground state energy
obtained by CF diagonalization.  In the spherical geometry, 
$E^{(J)}(L)$ and $\chi^{(J)}(L)$ will refer to the ground state energy 
and its wave function obtained by CF diagonalization 
in the orbital angular momentum $L$ sector.
As with all variational wave functions the energy $E^{(J)}$ 
is a strict upper bound on the exact energy.

Of course, by increasing $J$, one will eventually obtain the 
exact state (although through an unnecessarily complicated route).
The great advantage of the CF theory is that a very good approximation 
to the exact state is obtained by diagonalizing in a very small 
basis, in addition to providing an understanding of the physics. 
For many total angular momenta $L$ a unique state is obtained at the lowest 
level (i.e. 
$\Phi^{(0)}$ is unique), which gives
a variational state with no free parameters.
In the spherical geometry used in Sec.~\ref{gapsection} this happens 
only when $\nu^\ast$ is an integer.
On the disk there are many such states (c.f.\ Table \ref{results}), labeled 
compact states\cite{JainKawamura}.

The edge effects are significant in the disk geometry used in this section, 
while the spherical geometry used in Sec.~\ref{gapsection} represents the bulk.
In the spherical geometry, excitations are obtained by promoting a CF from the highest filled 
$\Lambda$L to the lowest empty $\Lambda$L, leaving behind a hole.
Thus in $\Psi_{\eta}^{(J)}$, $J$ CF particle-hole pairs exist.
We will find the $J$th order variational state in the subspace spanned by CF states 
with $\le J$ holes in the highest filled CF \LL\ of $\Psi^{(0)}$.
In the disk geometry, once a CF is raised to a higher $\Lambda$L, 
its angular momentum is no longer fixed, and one can redistribute the angular momenta
of the particles in the lowest $\Lambda$L.  This additional variational 
freedom lets us investigate edge effects as well.\cite{Jeon,Mandal1}

\begin{figure}[htbp]
\includegraphics[scale=0.65]{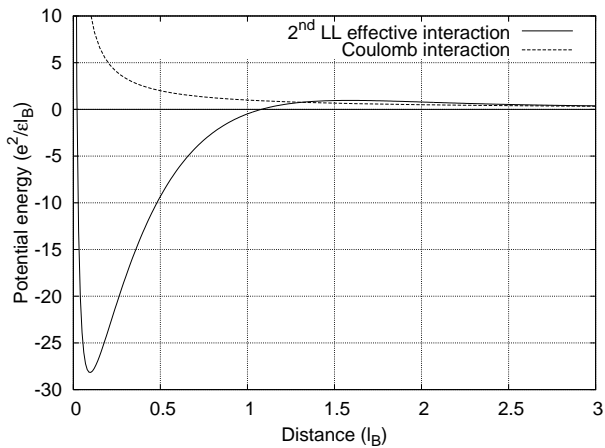}
\caption{\label{poten} The effective real-space 
potential $V^{\text{eff}}(r)$ (defined in Eq.~\ref{form}) 
which simulates the second LL physics in the 
lowest LL.  In spite of the deep dip at $r<l_B$, where $l_B$ is the 
magnetic length, it reproduces the 
\second\ LL pseudopotentials to a very high accuracy (details in text).
}
\end{figure}

Once the CF basis is constructed, one evaluates the overlaps 
$\langle\Psi^{(j_1)}_\alpha|\Psi^{(j_2)}_\beta\rangle$ between 
the basis vectors and the matrix elements of the
interaction $\langle\Psi^{(j_1)}_\alpha|H_I|\Psi^{(j_2)}_\beta\rangle$ for 
$j_1,j_2\le J$ in the $J$th order using a Metropolis Monte Carlo technique.
That requires the knowledge of the real-space interaction corresponding to 
the effective pseudopotentials $V^n_m$.
The inverse Fourier transform of 
$\tilde V^n(q)$ in Eq.~(\ref{effective}) is not defined, as the required integral is divergent.
This, however, causes no problem.  Because the pseudopotentials completely determine the 
interaction, any real-space interaction $V^n(r)$ that gives the same $V^n_m$ through
\begin{equation}
\label{pseudoeff}
V^n_m=\frac{\langle\psi_m|\sum_{i<j}V^n(z_i-z_j)|\psi_m\rangle}{\langle\psi_m|\psi_m\rangle},
\end{equation}
will be sufficient for our purposes.  
Thus we can use some convenient prescription for the analytic form of $V^n(r)$.
The goal is that the Monte Carlo evaluation of the matrix elements of $V^n(r)$, 
required by the diagonalization in the correlated CF basis,
should converge rapidly.  In this paper we will use the following form:
\begin{equation}
V^{\text{eff}}(r)=\frac{1}{r}+\sum_{i=0}^M c_i r^i e^{-r}.
\label{form}
\end{equation}
The form of the real-space interaction in Eq.~(\ref{form}) is based on the 
observation that the long-range behavior of the 
effective interaction should approach the Coulomb interaction, hence
all corrections must be short-range. 
Keeping enough number of terms in the sum will give as accurate a 
representation of the interaction as desired.   
For our purposes, it is enough to keep the first seven terms ($M=6$) in the sum.
We calculate the first seven odd $m$ pseudopotentials
$V^{\text{eff}}_1,V^{\text{eff}}_3,\dots,V^{\text{eff}}_{13}$ from Eq.~(\ref{form}) 
symbolically by Mathematica, and determine $c_i$ ($0\le i\le M$) to satisfy
\[V^{\text{eff}}_m=V^1_m\;.
\]
The coefficients $c_i$ thus obtained, and used in our calculations below, 
are given in Table \ref{coeff}, and the resulting 
real space interaction is shown in Fig.~\ref{poten}.
Alternative prescriptions are available in the literature.
Refs.\ \onlinecite{Lee1,Park2,Scarola2} use a Gaussian functional form 
for the same purpose:
\[
\frac{1}{r}+\sum_{i=0}^M c_i r^{2i} e^{-r^2}.
\]
This form, however, is less convenient for our purposes, because it 
yields a real-space potential that oscillates strongly as a function of distance,
causing an extremely slow convergence in Monte Carlo.

\begin{table}
\caption{\label{coeff} Coefficients  in 
%of the effective real-space 
%interaction $V^{\text{eff}}(r)$ of the second LL approximated in the form in 
Eq.~(\ref{form}) to produce the effective interaction for the second LL.}
%when fitting $V^1_1,\dots,V^1_{13}$.}
\begin{ruledtabular}
\begin{tabular}{rl}
Coefficient & Value \\ \hline
$c_0$ & -50.36588 \\
$c_1$ & 87.38159 \\
$c_2$ & -56.08439 \\
$c_3$ & 17.76573 \\
$c_4$ & -2.97162 \\
$c_5$ & 0.25132 \\
$c_6$ & -0.008435
\end{tabular}
\end{ruledtabular}
\end{table}

We comment on the validity of the truncated interaction, i.e. 
Eq.~(\ref{form}) with $M=6$ and the coefficients in Table \ref{coeff}.
First, the physics of the FQHE is governed by the first few relevant 
pseudopotentials.  Since our truncated interaction reproduces correctly, by 
construction, the first {\em seven} odd pseudopotentials, it is expected to be quite good. 
Second, $V^{\text{eff}}$ is also guaranteed to reproduce the pseudopotentials 
for very large $m$, where the interaction approaches $1/r$.  We have found 
that the largest relative error in $V_m$ is 5\% at $m=37$ up to $m\leq 49$.  
(For a given total angular momentum $L$, the greatest relative 
angular momentum is $m_{max}=L-\binom{N-1}{2}$; for our calculations, which go 
up to $L=60$ for $N=6$, the largest relevant relative angular momentum is 49.)
At such large $m$, the pseudopotential is so small that such error is of 
no consequence.  As a final test, we have compared the ground state of 
the truncated interaction with the ground state of the Coulomb interaction. 
In the $L\le 75$ range the smallest overlap occurs at $L=71$, which is 0.998303.
To quote a typical number, at $L=45$ the overlap is 0.99997.
The truncated interaction is thus essentially exact.

The overlap and interaction matrix elements for the effective interaction 
were calculated by the Metropolis 
algorithm. Typically $6.6\times 10^{7}$ to $7.5\times 10^{7}$ Monte Carlo steps (MCS) were performed,
where a MCS is defined as a number of iterations during which each particle 
coordinate is expected to be updated once.
Averages and error bars were calculated from five independent runs.
When a sufficient accuracy was reached, the standard Gram-Schmidt 
procedure was used to find an orthonormal basis (often telling us that not all of the vectors
in the correlated CF basis were independent), and $H_I$ was diagonalized.

\section{Results}
\label{resultssec}

\begin{figure}
\includegraphics[scale=0.65]{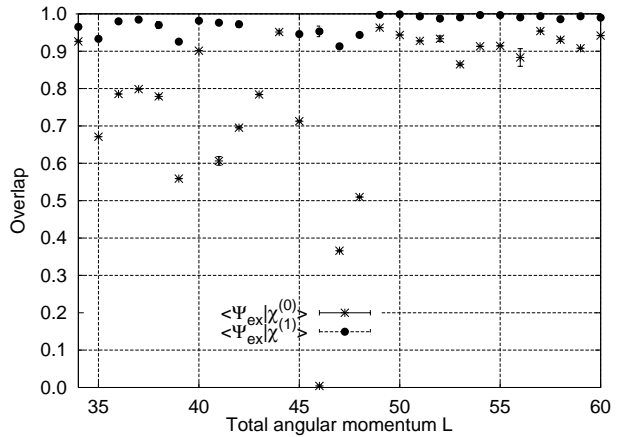}
\caption{\label{overlap} Overlap of the exact \second\ Landau level ground
state with the wave function obtained from CF diagonalization with and without 
\LL\ mixing (solid circles and stars, respectively).}
\end{figure}

We have performed CF diagonalization in the $34\le L\le 60$ range. 
This encompasses $\nu=1/3$ (which occurs at $L=45$).
The flavors $^2\text{CF}$'s are used for $L\le 48$, $^4\text{CF}$'s for $L\ge 49$.
(This choice ensures the smallest CF basis.)
The results are shown in Table \ref{results} and Fig.~\ref{overlap}.
The zeroth-order variational energies and overlaps are given for all $34\le L\le 60$. 
The first-order wave function is evaluated for all $L$'s except for
$L=43,44$; for these cases the correlated CF basis is too big ($D^{(1)}=83$ and 111, 
respectively) preventing an evaluation of the matrix elements of $H_I$ with  
sufficient accuracy.

\begin{table}
\caption{\label{results}
Comparison of the energies (in units of $e^2/\epsilon l_B$) of the exact ground
state and the states obtained from \zeroth\ and \first\ order CF diagonalization\cite{ourfootnote}.
The number in parentheses is the error in the last digit.
$D_{ex}$, $D^{(0)}_{CF}$ and $D^{(1)}_{CF}$ are the dimensions of the full Hilbert 
space, of the CF basis in the \zeroth\ order calculation, and of the CF basis 
in the \first\ order calculations, respectively.
The \first\ order energy was not calculated for $L=43,44$ because $D^{(1)}_{CF}$ is 
very large, which makes the numerical calculation very time consuming.}
\begin{ruledtabular}
\begin{tabular}{rrrrrrr}
$L$ & $E_{ex}$ & $E^{(0)}_{CF}$ & $E^{(1)}_{CF}$ & $D_{ex}$ & $D^{(0)}_{CF}$ &
$D^{(1)}_{CF}$ \\ \hline
34 & 3.57207 & 3.5801(1) & 3.5748(7) & 235 & 4 & 27 \\
35 & 3.48739 & 3.5144(1) & 3.49786(2) & 282 & 1 & 10 \\
36 & 3.44351 & 3.47805(2) & 3.4476(2) & 331 & 2 & 18 \\
37 & 3.39199 & 3.42079(1) & 3.3962(1) & 391 & 5 & 31 \\
38 & 3.37222 & 3.3899(3) & 3.3770(5) & 454 & 9 & 47 \\
39 & 3.31020 & 3.3403(1) & 3.3174(3) & 532 & 1 & 17 \\
40 & 3.25712 & 3.27476(2) & 3.2616(1) & 612 & 2 & 26 \\
41 & 3.22851 & 3.2647(3) & 3.2330(5) & 709 & 4 & 41 \\
42 & 3.18755 & 3.2206(3) & 3.1933(9) & 811 & 7 & 59 \\
43 & 3.15253 & 3.1761(1) & - & 931 & 12 & 83 \\
44 & 3.12831 & 3.1379(4) & - & 1057 & 18 & 111 \\
45 & 3.05354 & 3.0796(3) & 3.0621(1) & 1206 & 1 & 28 \\
46 & 3.04126 & 3.0805(5) & 3.0484(6) & 1360 & 1 & 39 \\
47 & 3.01108 & 3.0561(3) & 3.0235(5) & 1540 & 2 & 55 \\
48 & 2.97252 & 3.02019(8) & 2.9860(8) & 1729 & 3 & 74 \\
49 & 2.94205 & 2.94937(4) & 2.9443(2) & 1945 & 4 & 46 \\
50 & 2.87713 & 2.88772(2) & 2.8805(2) & 2172 & 2 & 32 \\
51 & 2.87227 & 2.8786(6) & 2.8790(2) & 2432 & 1 & 19 \\
52 & 2.85393 & 2.8667(3) & 2.8597(6) & 2702 & 10 & 65 \\
53 & 2.82087 & 2.8375(3) & 2.827(1) & 3009 & 5 & 44 \\
54 & 2.77835 & 2.79440(4) & 2.7881(5) & 3331 & 2 & 26 \\
55 & 2.72364 & 2.737(2) & 2.7289(2) & 3692 & 1 & 13 \\
56 & 2.72364 & 2.7457(2) & 2.7383(2) & 4070 & 5 & 39 \\
57 & 2.69540 & 2.70221(2) & 2.7018(1) & 4494 & 2 & 21 \\
58 & 2.68363 & 2.7054(7) & 2.6979(7) & 4935 & 9 & 48 \\
59 & 2.64022 & 2.65914(2) & 2.6495(6) & 5427 & 3 & 25 \\
60 & 2.58918 & 2.603(7) & 2.6034(4) & 5942 & 1 & 9
\end{tabular}
\end{ruledtabular}
\end{table}

The principal result of our study, as seen in 
Table \ref{results} and Fig.~\ref{overlap}, is that 
the CF theory without $\Lambda$L mixing is not satisfactory, but 
very good agreement is obtained once $\Lambda$L mixing is incorporated 
at the lowest order.  Let us consider a few specific cases.

At $\nu=\myfrac{1}{3}$ ($L=45$) the zeroth-order variational wave function 
$\chi^{(0)}$ is the Laughlin wave function, the energy of which 
($E_{CF}^{(0)}=3.0796(3)$) significantly overestimates the exact energy 3.05354.
The overlap is 0.712(2), which is similar to the values that were found in 
the spherical geometry \cite{Ambru}.  The first-order CF diagonalization 
yields $E_{CF}^{(1)}=3.0621(1)$ and $|\langle\Psi_{ex}|\chi^{(1)}\rangle|=0.9467(8)$.

We commented earlier that for certain $L$ values the ground states 
of different LL's have anomalously low overlaps.
Let us consider $L=41$, 46, and 52 for $N=6$.  (We could not investigate 
$L=81$ due to computational limitations.) 
At the zeroth order (i.e., without $\Lambda$L mixing), the overlaps 
for these $L$ are $|\langle\Psi_{ex}|\chi^{(0)}\rangle|=0.614(2)$, 
0.004(2) and 0.933(7), respectively.
For $L=46$ the small overlap follows because the \zeroth\ order CF basis contains only 
one state, which is very close to the LLL ground state.  In the other 
two cases, the \zeroth\ order CF basis is larger (with 4 and 10 states); the greater 
flexibility results in better overlaps.  With $\Lambda$L mixing, the overlaps 
increase to $|\langle\Psi_{ex}|\chi^{(1)}\rangle|=0.9814(2)$, 0.9695(7) and 0.987(7), 
respectively.

%The angular momentum $L=56$ for $N=6$ deserves further 
%comment.  Here, the exact ground 
%states of the lowest two Landau levels are very nearly identical, but neither 
%is well captured by the variational ground state of the lowest 
%order CF diagonalization.  Here, $\Lambda$L mixing is important even 
%in the lowest Landau level.  It is obviously a finite size effect, which 
%we have not investigated further. 

\section{Estimation of the Gap at $\nu^{(1)}=1/3$}
\label{gapsection}

As was mentioned above, the existence of an excitation gap is crucial
for the theoretical explanation of the phenomena of the FQHE.
The best estimate so far is from the exact diagonalization calculation 
of Morf~\cite{Morf1}, where he obtained a 
gap of $\Delta_{7/3}\approx 0.02 e^{2}/\epsilon l_B$
for $N=10$ particles at $\nu^{(1)}=1/3$, which gives a measure of the 
thermodynamic gap.  In this section we estimate the excitation gap 
at $\nu^{(1)}=1/3$ from CF diagonalization, incorporating
the residual interactions among CF's.

For the calculation of excitation gaps we switch to the spherical geometry,
which is good for examining the bulk properties.  In this  
geometry, electrons are constrained to move on the
surface of a sphere and a radial magnetic field is produced by
placing a magnetic monopole of strength $Q$ at the center, where
$2Q\phi_0$ is the magnetic flux through the surface of the sphere
($2Q$ is an integer according to Dirac's quantization condition.)

On the sphere the total {\em orbital} angular momentum $L$ 
is a good quantum number.  The  ground state at $L=0$
is uniform and spherically symmetric and is considered incompressible
if it is separated by a finite excitation gap from the other states.

The single particle states $Y_{qlm}$ are called monopole  harmonics
\cite{Yang1976} and are given by
\begin{eqnarray}
Y_{qlm}(\Omega) &=& N_{qlm}(-1)^{l+m}e^{iq\phi}u^{q-m}v^{q+m}\\
\nonumber &&\times \sum_{s=0}^{l-q} \binom{l-q}{s}\binom{q+l}{l+m-s}\\
\nonumber  &&\times (v^{*}v)^{l-q-s}(u^{*}u)^s \;,
\end{eqnarray}
where $q$ is the monopole strength, $l=q+n$ is  the single particle
angular momentum where $n=0,1,\ldots$ is the LL index,
$m=-l,-l+1,\ldots,l$ is the z-component of angular  momentum,
$\Omega=(\theta,\phi)$ is the position of the electron  on the surface
of the sphere in the usual coordinates,  and
$u\equiv\cos(\theta/2)\exp (-i\phi /2)$ and $v\equiv\sin(\theta/2)\exp
(i\phi /2)$.   The normalization coefficient is
\begin{eqnarray}
N_{qlm} = \sqrt{\frac{2l+1}{4\pi}\frac{(l+m)!(l-m)!}{(l+q)!(l-q)!}}\;.
\end{eqnarray}
The degeneracy of the $n$th $\Lambda$-level is $2l+1$ and the  filling
factor is defined as
\begin{eqnarray}
\label{sphericalnu}
\nu = \lim_{N\rightarrow \infty}\frac{N}{2Q} \;.
\end{eqnarray}
The distance between particles $r_{ij}$ is taken to be the cord
distance $r_{ij} = 2R|u_i v_j - v_i u_j|$ where the radius of the
sphere is $R=\sqrt{Q}$ in units of magnetic length.

In the spherical geometry the CF wave function is
written in the same form as in Eq.~(\ref{mapping})
\begin{eqnarray}
\Psi = P_{LLL}\Phi_1^{2p}\Phi \nonumber \;,
\end{eqnarray}
where $\Phi$ is now a wave function for $N$ non-interacting electrons
at monopole strength $q$.  Composite fermionization 
(vortex attachment) is again 
accomplished by multiplication of the Jastrow factor $\Phi_1^{2p}$ which
in this geometry is written
\begin{eqnarray}
\label{Jsphere}
\Phi_1^{2p} = \prod_{j<k} (u_j v_k - v_j u_k)^{2p} 
\end{eqnarray}
Again, $\Phi_1$ is the wave function of one filled LL.
The projection $P_{LLL}$ into the lowest LL in the spherical geometry 
is a complicated procedure and interested reader is
again referred to the literature \cite{JainKamilla}.
We then arrive at a wave function $\Psi$ describing $N$ electrons at $Q = q + p(N-1)$.

To obtain states  corresponding to the filling 
factors ($\nu=n/(2pn+1)$) we create electron
states at integral filling factors $n$.   In the spherical geometry
this amounts to setting the monopole strength to $2q=N/n-n$, the value
corresponding to a total degeneracy of $N$.  After attaching $2p$ vortices
to each electron and projecting into the LLL we arrive at a wave
function  describing $N$ interacting electrons at filling factor
$\nu=n/(2pn+1)$ called $\Psi^{(0)}$ (this wave function equivalently 
describes non-interacting CFs).  As in the disk geometry $\Psi^{(0)}$ 
has been shown to be
spectacularly accurate when compared with exact diagonalization results
\cite{JainKamilla}.  On the sphere $\Psi^{(0)}$ is a uniform state and exists
only at $L=0$.  With no loss of generality  we work within the
$z$-component of angular momentum $L_z=0$ subspace of the $2L+1$
degenerate states in each $L$ channel.

The low energy excited states $\Psi_{\alpha}^{(1)}$ are calculated as
\begin{eqnarray}
\Psi_{\alpha}^{(1)} =
P_{LLL}\Phi_1^{2p}\Phi_{\alpha}^{(1)}\;,
\end{eqnarray}
where $\Phi_{\alpha}^{(1)}$ represents  the $\alpha$ states at
$2q=N/n-n$ where one electron in the filled LL $n$ has been
allowed to occupy the LL $(n+1)$ leaving behind a hole.  In
general $\Phi_{\alpha}^{(1)}$  is a superposition  of Slater
determinants.  There is one state for each total angular momentum
$L=1,2,\ldots,(N+n^2+n)/n$, so no CF diagonalization is required, 
though it has been shown \cite{L1annihilate} that the application 
of $P_{LLL}$ annihilates  the
state at $L=1$.

\begin{figure}
\includegraphics[width=3.25in,angle=0]{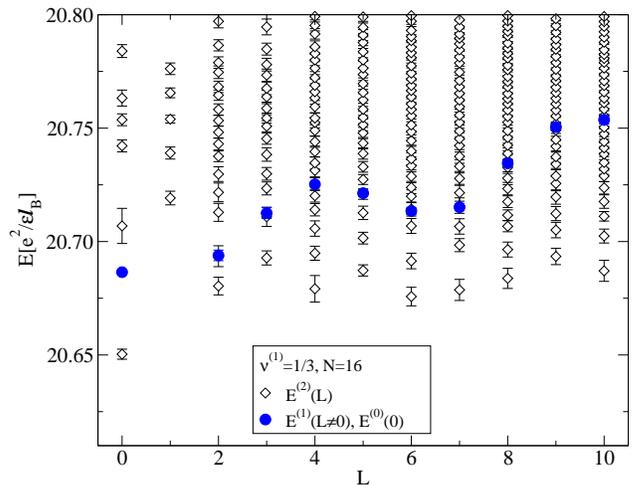}
\caption{\label{J1vsJ2} (Color online) The open diamonds
represent the energy spectra for the $J=2$ basis as a function
of total orbital angular momentum $L$, i.e. $E^{(2)}(L)$, while 
the blue circles represent the 
energy spectra using the $J=1$ basis, i.e. $E^{(1)}(L\neq 0)$,  at filling
factor $\nu^{(1)}=1/3$ for $N=16$ electrons.  The energies
have been corrected for the finite sized deviation in the density by  
multiplication by $\sqrt{\rho /\rho_N}=\sqrt{2Q\nu/N}$.  
The contribution from the uniform positive
background has not been subtracted.
}
\end{figure}

The calculated excitation spectrum $E^{(1)}(L)$ for $\nu^{(1)}=1/3$ for 
$N=16$ is shown by solid circles in Fig.~\ref{J1vsJ2}.  (We
are showing a system for $N=16$ for illustrative purposes only; 
similar behavior occurs for other system sizes.)  It has  
an $L=0$  ground state ($\Psi^{(0)}$ in this case)
separted from the other states by a gap.  It may be tempting to go
ahead and calculate the excitation gap, but we know that
$\Psi^{(0)}$ is a poor representation of the actual ground 
state in the second LL from the studies of small systems in the 
spherical geometry \cite{Ambru} and in the disk geometry (this
work and elsewhere \cite{JainKamilla}) and suspect that the excited
states are also equally bad.  Thus, we expand our 
basis for CF diagonalization (Eq.~\ref{basis}) to $J=2$.

The open diamonds in Fig.~\ref{J1vsJ2} show the spectra when the
interaction  in Eq.~(\ref{form}) is diagonalized using the CF basis in
Eq.~(\ref{basis}) with $J=2$ for $N=16$ at $\nu^{(1)}=1/3$.   Two 
points are noteworthy.  First, as expected, the lowest energy in each 
$L$ sector has been lowered significantly.
(Many new higher energy states are also created, but they
are not of interest in this work.) Second, the gap is significantly 
enhanced.  Both these underscore the importance of $\Lambda$L mixing 
for the second LL FQHE. 
Table \ref{overlaps} gives the overlap between
$\chi^{(2)}$ at  $L=0$ and $\Psi^{(0)}$ for a number of system sizes.  
The relatively small overlaps give further evidence that 
the CF diagonalization in the $J=2$
basis has significantly improved the ground state wave function.

The overall size of the  basis
diagonalized for $J=1$ consists of $(N+n^2+n)/n+1$ states across the
whole spectrum while the size of the basis diagonalized for $J=2$ is
approximately $500$ states for $N=16$ ($9$ states in the $L=0$ channel
up to $\sim 50$ states in the $L=10$ channel.)   That should be
compared to the dimension of the subspace for the exact state which
is $\sim 10^{10}$,  making it inaccessible to exact diagonalization.
The overlap and interaction matrix elements needed for CF 
diagonalization used on average $10^7$
Monte Carlo iterations.  The error reported in each value is the
standard deviation of the mean calculated between, on average, ten
separate Monte Carlo configurations.  Running in parallel  on a
multi-node computer cluster, approximately $7200$ hours of computer
time  were utilized in the spherical geometry calculations.  To obtain
the final  energy spectra we implement the Gram-Schmidt procedure to
diagonalize  the interaction Hamiltonian within the subspace spanned
by the $J=2$ basis as it is generally not orthogonal.

A technical detail regarding the sampling function is noteworthy.  
We choose this function to be one of the basis states that we
feel closely represents the true state.   In principle, in the limit
of infinite Monte Carlo iterations the result calculated approaches
the exact result regardless of sampling function.  In practice, a proper  
choice of sampling function can make the Monte  Carlo converge faster.  Using
the effective potential in Eq.~(\ref{form})  we have found that the Monte
Carlo error is on the order of $10$ times larger than the error when using the
pure Coulomb potential when $\Psi^{(0)}$ is used as the sampling
function.  Consequently, we have calculated the spectra using many different
sampling functions.  Since we are only interested in the lowest energy in
each $L$ channel, this energy was chosen to be the energy  with the
smallest statistical error among the energies calculated with the
different sampling functions.

\begin{table}
\caption{\label{overlaps} The overlap between the improved variational 
ground state wave function and the Laughlin wave function at $\nu^{(1)}=1/3$
($\langle\chi^{(2)}|\Psi^{(0)}\rangle$)
for $N=10,11,12,13,$ and $16$.  The number given in parentheses
is the Monte Carlo uncertainty.}
\begin{ruledtabular}
\begin{tabular}{rl}
$N$ & $|\langle\chi^{(2)}|\Psi^{(0)}\rangle|^2$ \\ \hline 
$10$ & 0.82(1) \\  $11$ & 0.81(1) \\    $12$ & 0.82(2) \\  $13$ & 0.835(3) \\
$16$ & 0.77(4)
\end{tabular}
\end{ruledtabular}
\end{table}

\begin{figure}
\includegraphics[width=3.25in,angle=0]{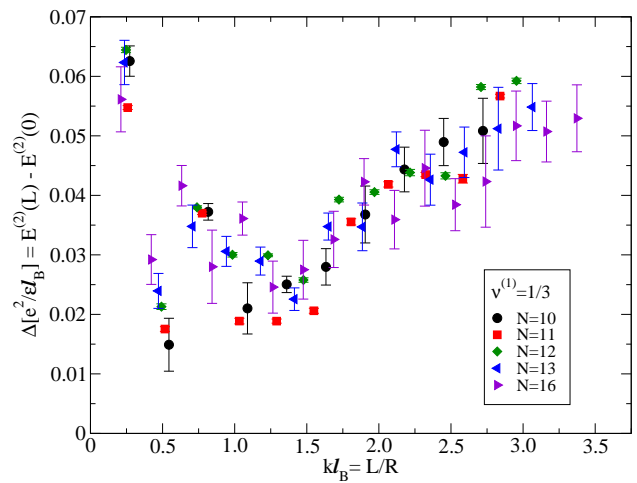}
\caption{\label{gap} (Color online) The different symbols represent
the energy gap 
in units of $e^{2}/\epsilon l_B$ for systems sizes $N=10,11,12,13$, and $16$ 
as a function of $k l_B = L/R$, where $R$ is the radius 
of the sphere and $L$ is the total orbital angular momentum.
Each energy  has been corrected for finite sized deviation in the
density as in Fig.~\ref{J1vsJ2}.}
\end{figure}

We believe that $J=2$ CF diagonalization is quantitatively accurate 
in the second LL, and proceed to calculate the excitation gap for 
FQHE at $\nu^{(1)}=1/3$.
Fig.~\ref{gap} shows the energy  gap $\Delta$ as a function of wave
vector $k$, where $kl_B=L/R$.   We have calculated the energy gap for
$N=10,11,12,13,$ and $16$ represented  by the different symbols.  The
energies for different system sizes have roughly collapsed onto a single
line indicating that we are close to the thermodynamic limit.  The
Monte Carlo error  is $\sim 10$ times what one is used to with gap
calculations in the lowest LL, making the numbers less certain.
However, many conclusions can still be reached.

The shape of $\Delta$ is somewhat different than that of the lowest
LL $1/3$ state which shows one roton minimum.  There
seem to be two minima in the spectra, at $kl_B\sim 0.5$ and
$kl_B\sim 1.25$.  The value of roton gap at
both $kl_B\sim 0.5$ and $kl_B\sim 1.25$ is crudely $\sim 0.02
e^{2}/\epsilon l_B$ in both cases.
This is consistent with Morf's estimation\cite{Morf1}, 
further confirming that the CF theory with lowest order $\Lambda$L mixing
is quite satisfactory for FQHE in the second LL.
The transport gap is identified with the excitation energy at $L=N$ since in the 
thermodynamic limit this
state describes an infinitely separated CF particle-hole pair; this 
is estimated to be $\sim 0.05 e^{2}/\epsilon l_B$.
The theoretical gaps are much larger than the gaps measured 
experimentally, however. 
In Ref.\ \onlinecite{2ndLLGaps} the excitation gaps
at $7/3$  and $8/3$ were measured as $\Delta_{7/3} = 0.10 $ K  and
$\Delta_{8/3} = 0.055 $ K, respectively.  Using the  experimental
parameters our calculated gap for $\nu=7/3$ is
$\Delta \sim 2 $ K for the roton and $\Delta \sim 5 $ K
for the transport  gap.  The 
theoretical calculations quite generally overestimate the FQHE gap; we expect that 
our estimate would scale down once finite thickness, LL mixing, and disorder 
effects are taken into account.  In this context, it may be noted
that the theoretical gap at $\nu=5/2$ 
estimated from exact diagonalization studies by Morf\cite{Morf1},
$\Delta_{5/2}\approx 0.05 e^2/\epsilon l_B$, corresponding to $\sim 5$ K,
is also large compared to the experimental gap\cite{2ndLLGaps} ($\sim 0.11$ K).
%It is also worth mentioning that our gap is significantly lower than 
%$0.15 e^2/\epsilon l_B$ quoted in Ref.~\onlinecite{Goerbig04b}.

In principle, our method allows one to
obtain the excitation gap for fillings of the form $\nu^{(1)}=n/(2pn+1)$.  
However, in practice we find that the Monte Carlo error is too large
for any reasonable number of iterations,
perhaps due to the nature of the effective  potential (Eq.~\ref{form}) 
at short distances and the smallness of the gap.

\section{Conclusion}

We have shown that the residual interaction between CF's accounts
for the deviation of \second\ LL FQHE states 
and the LLL FQHE states in certain range of filling factors. 
This interaction can be taken into
account by allowing for \LL\ mixing, i.e., letting the ground state hybridize 
with CF particle-hole excitations across \LL's, which is caused by 
the inter-CF interactions.  The lowest level incorporation of $\Lambda$L 
mixing produces an excellent account of the second LL ground states. 
We also estimate the excitation gap at $\nu^{(1)}=1/3$ 
and find that $\Lambda$L mixing strongly renormalizes it upwards.

\section*{Acknowledgments}

We are indebted to Chia-Chen Chang for useful discussions and for sharing with 
us several numerical codes he developed in other projects.  We are grateful 
to the High Performance Computing (HPC) group at Penn State University ASET 
(Academic Services and Emerging Technologies)
for assistance and computing time on the Lion-XL and Lion-XO clusters.  Partial 
support of this research by the National Science Foundation under 
grants No.\ DMR-0240458 and DGE-9987589 (IGERT) is gratefully acknowledged.

\newcommand{\journal}[1]{{#1}}
\newcommand{\PRL}{\journal{Phys.\ Rev.\ Lett.}}
\newcommand{\PRB}{\journal{Phys.\ Rev.\ B}}

\end{document}